\documentclass[10pt,twoside,twocolumn,english,aps,prl,superscriptaddress,notitlepage]{revtex4-2}
\usepackage[utf8]{inputenc}
\usepackage{afterpage}
\usepackage[a4paper]{geometry}
\usepackage{color,xcolor,soul}
\usepackage{verbatim,textcomp}
\usepackage{amstext,amsfonts,amsmath,amssymb,pifont}
\usepackage{graphicx}
\usepackage[calcwidth,explicit]{titlesec}
\usepackage[normalem]{ulem}
\usepackage[unicode=true,pdfusetitle, bookmarks=false, breaklinks=true,pdfborder={0 0 0},pdfborderstyle={},backref=false,colorlinks=true]{hyperref}
\usepackage{array}
\usepackage{multirow}
\usepackage{adjustbox}
\hypersetup{colorlinks=true,citecolor=black,linkcolor=black,urlcolor=black,pagebackref=true}
\newcolumntype{C}[1]{>{\centering\let\newline\\\arraybackslash\hspace{0pt}}m{#1}}

\makeatletter\renewcommand\frontmatter@abstractwidth{\dimexpr\textwidth-2cm\relax}\makeatother

\titleformat{\section}{\bfseries\sffamily\large\filcenter}{\thesection.}{0.2em}{#1}
\titlespacing{\section}{0pt}{0.2ex}{0.2ex}
\titleformat{\paragraph}[runin]{\normalfont\normalsize\bfseries}{}{0pt}{\theparagraph.}
\titlespacing*{\paragraph}{0em}{0ex}{0.3em}[]

{}

\setcitestyle{super}

\renewcommand\thesection{\Alph{section}}
\renewcommand{\theparagraph}{\Alph{section}\arabic{paragraph}}
\makeatletter\@addtoreset{paragraph}{section}\makeatother
\makeatletter\def\p@paragraph{}\makeatother

\AtBeginDocument{
\renewcommand{\ref}[1]{\autoref{#1}}

\newcommand\fakesubsection[1]{
    \refstepcounter{subsection}  
    \addcontentsline{toc}{subsection}{\protect\numberline{\thesubsection}#1} 
}
\newcommand{\fakesection}[1]{
    \par\refstepcounter{section}
    \sectionmark{#1}
    \addcontentsline{toc}{section}{\protect\numberline{\thesection}#1}
}
}

\begin{document}
\title{Visualizing the Strongly Reshaped Skyrmion Hall Effect \\in Multilayer Wire Devices\smallskip{}}
\author{Anthony K.C. Tan}
\thanks{These authors contributed equally to this work}
\affiliation{Data Storage Institute, Agency for Science, Technology \& Research (A{*}STAR), Singapore}
\affiliation{Cavendish Laboratory, University of Cambridge, Cambridge, UK}
\author{Pin Ho}
\thanks{These authors contributed equally to this work}
\email{hopin@imre.a-star.edu.sg}
\affiliation{Institute of Materials Research \& Engineering, Agency for Science, Technology \& Research (A{*}STAR), Singapore}
\affiliation{Data Storage Institute, Agency for Science, Technology \& Research (A{*}STAR), Singapore}
\author{James Lourembam}
\affiliation{Institute of Materials Research \& Engineering, Agency for Science, Technology \& Research (A{*}STAR), Singapore}
\affiliation{Data Storage Institute, Agency for Science, Technology \& Research (A{*}STAR), Singapore}
\author{Lisen Huang}
\affiliation{Institute of Materials Research \& Engineering, Agency for Science, Technology \& Research (A{*}STAR), Singapore}
\affiliation{Data Storage Institute, Agency for Science, Technology \& Research (A{*}STAR), Singapore}
\author{Hang Khume Tan}
\affiliation{Institute of Materials Research \& Engineering, Agency for Science, Technology \& Research (A{*}STAR), Singapore}
\affiliation{Data Storage Institute, Agency for Science, Technology \& Research (A{*}STAR), Singapore}
\author{Cynthia J.O. Reichhardt}
\affiliation{Theoretical Division and Center for Nonlinear Studies, Los Alamos National Laboratory, Los Alamos, NM, USA}
\author{Charles Reichhardt}
\affiliation{Theoretical Division and Center for Nonlinear Studies, Los Alamos National Laboratory, Los Alamos, NM, USA}
\author{Anjan Soumyanarayanan}
\email{anjan@imre.a-star.edu.sg}
\affiliation{Institute of Materials Research \& Engineering, Agency for Science, Technology \& Research (A{*}STAR), Singapore}
\affiliation{Data Storage Institute, Agency for Science, Technology \& Research (A{*}STAR), Singapore}
\affiliation{Physics Department, National University of Singapore, Singapore}

\begin{abstract}
\noindent Magnetic skyrmions are nanoscale spin textures touted as next-generation computing elements. When subjected to lateral currents, skyrmions move at considerable speeds. Their topological charge results in an additional transverse deflection known as the skyrmion Hall effect (SkHE). While promising, their dynamic phenomenology with current, skyrmion size, geometric effects and disorder remain to be established. Here we report on the ensemble dynamics of individual skyrmions forming dense arrays in Pt/Co/MgO wires by examining over 20,000 instances of motion across currents and fields. The skyrmion speed reaches 24~m/s in the plastic flow regime and is surprisingly robust to positional and size variations. Meanwhile, the SkHE saturates at $\sim 22^\circ$, is substantially reshaped by the wire edge, and crucially increases weakly with skyrmion size. Particle model simulations suggest that the SkHE size dependence -- contrary to analytical predictions --- arises from the interplay of intrinsic and pinning-driven effects. These results establish a robust framework to harness SkHE and achieve high-throughput skyrmion motion in wire devices.
\end{abstract}
\maketitle

\section*{Introduction}

\fakesubsection{Sk-Motion Motivation}
Magnetic skyrmions are nanoscale, topologically wound spin structures stabilized in a ferromagnetic background by competing magnetic interactions\citep{Nagaosa2013,Wiesendanger2016}. Their discovery at room temperature (RT) in chiral multilayer films comprising heavy metal -- ferromagnet interfaces has sparked scientific and technological excitement\citep{Boulle2016,MoreauLuchaire2016,Woo2016,Soumyanarayanan2017}. Notably, the spin-orbit torque (SOT) generated at such interfaces by an in-plane charge current\citep{Miron2010a,Miron2011}  provides the ideal instrument for electrical manipulation of skyrmions\citep{Sampaio2013,Jiang2015,Woo2016}. Consequently, several device proposals seek to harness current-driven skyrmion motion within a wire, or `racetrack' architecture\citep{Parkin2008,Fert2017} towards applications in analog memory\citep{Tomasello2014}, logic\citep{Luo2018}, and synaptic computing\citep{Li2017,Pinna2018}. In this light, achieving deterministic, efficient, and high throughput skyrmion motion in wires is a critical challenge for the device community.

\fakesubsection{Sk-Motion Literature}
Initial demonstrations of skyrmion motion were extremely promising -- showing efficient SOT manipulation\citep{Jiang2015} and individual speeds of up to $\sim 100$~m/s\citep{Woo2016}. However, progress since has been limited by significant challenges along three key fronts
pertaining to the influence of intrinsic, extrinsic, and collective effects. First, in addition to the expected linear motion, an applied current also induces a traverse skyrmion deflection -- known as the skyrmion Hall effect (SkHE) --  which arises from the hydrodynamic Magnus force acting on the skyrmionic topological charge\citep{Nagaosa2013,Jiang2016,Litzius2016,Reichhardt2016}. SkHE may enable defect avoidance\citep{Iwasaki2013b}, however the transverse deflection -- over 30$^\circ$ in some cases -- may limit linear mobility\citep{Litzius2016,Hrabec2017,Juge2019}. While efforts to develop SkHE-free materials are underway\citep{Woo2018,Dohi2019,Hirata2019,Zhang2016a,Zhang2016b}, several facets of SkHE remain unresolved amidst conflicting experimental results, e.g. the material-dependence of its magnitude, the existence of a saturation value, dependence on skyrmion size, etc.\citep{Jiang2016,Litzius2016,Woo2018,Juge2019,Zeissler2020}. Next, skyrmion dynamics in sputtered multilayer wires may be affected on one hand by material granularity and defects\citep{Kim2017,Legrand2017}, and on the other by interactions with the wire edge\citep{Iwasaki2013,Yoo2017}. However, these extrinsic effects are yet to be experimentally understood. Finally, high throughput devices would require skyrmion motion at densities $10-100$ times higher than prevailing experiments\citep{Parkin2008}, which have examined sparse configurations ($<1\,\mu$m$^{-2}$).

\fakesubsection{Results Summary}
Here we report on the ensemble dynamics of 80$-$200 nm sized skyrmions forming dense ($>10\,\mu$m$^{-2}$) arrays in Pt/Co/MgO multilayer wires. Using magnetic force microscopy (MFM) imaging, we examine over 20,000 instances of skyrmion motion over a range of applied currents and fields, spanning three distinct dynamic regimes: stochastic creep, deterministic creep and plastic flow. The onset of the deterministic motion is associated with finite SkHE, which grows and saturates at a moderate value ($\sim 22^\circ$). While the velocity is found to be surprisingly robust to edge effects and skyrmion size variations, the SkHE is considerably reshaped in both cases. Our simulations suggest that the observed SkHE trend with skyrmion size -- contrary to defect-free theoretical predictions -- arises from the interplay of intrinsic and pinning-driven effects. Our results and insights establish a robust experimental framework to realize high-throughput skyrmion motion in wire devices for next-generation nanoelectronics.

\section*{Results\label{sec:Wires-Setup}}

\begin{figure*}
\begin{centering}
\includegraphics[width=6in]{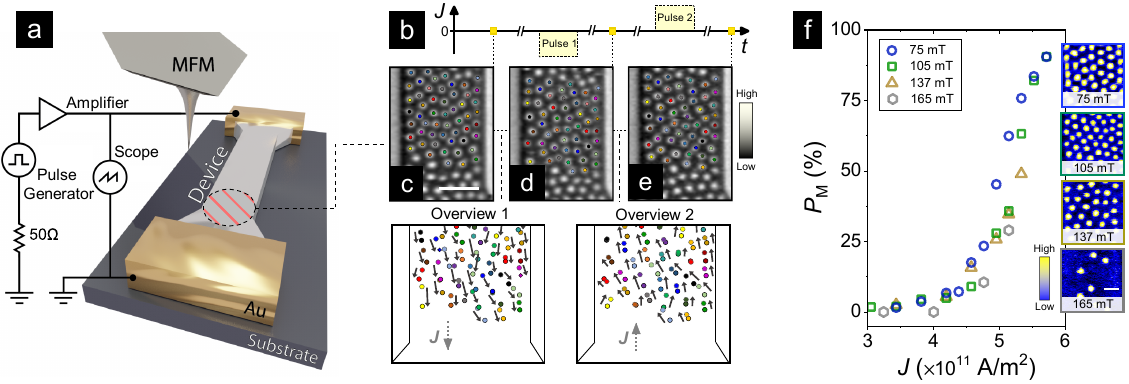}
\end{centering}
\caption[Skyrmion Motion Imaging Setup]
{\textbf{Experimental Setup and Device Characterization.}
\textbf{(a)} Schematic of the experimental setup. The [Pt(3)/Co(1.2)/MgO(1.5)]$_{15}$ (thicknesses in nm in parentheses) was mounted onto the MFM setup with varying \textit{in situ} out-of-plane (OP) magnetic fields (details in Methods). 
\textbf{(b)} Protocol used to inject current pulses (magnitude $J=\pm(1.0-5.8)\times10^{11}$~A/m$^2$, width: 20~ns) into the device. Pulses of alternating polarity were applied sequentially, and the wire was imaged by \emph{in situ} MFM before and after each pulse.
\textbf{(c--e)} Representative MFM images (scalebar: $1\,\mu$m) acquired at OP magnetic fields $\mu_{0}H\simeq105$~mT before pulsing (c), after applying current pulses of $J\simeq\mp 5.7\times10^{11}$~A/m$^2$ (d, e). Bottom left (right) panel shows tracked positions of selected skyrmions identified by coloured dots for $-J$ (c, d) and $+J$ (d, e) (details in SM4), whose motion is indicated by arrows.
\textbf{(f)} Proportion, $P_{\rm M}$, of skyrmions in motion along the driving current direction plotted as a function of $J$ for various OP fields. Insets show MFM images (scalebar: 500~nm) of the wire at varying $\mu_{0}H$ (magnitudes in insets) following the skyrmion nucleation recipe (see Methods). Finite $P_{\rm M}$ indicates the onset of deterministic skyrmion motion.\label{fig:SkMotion-Setup}}
\end{figure*}

\fakesubsection{Materials \& Experimental Setup}
\textbf{Imaging Skyrmion Dynamics.} This work was performed at RT on [Pt(3)/Co(1.2)/MgO(1.5)]$_{15}$ multilayer films sputtered on Si/SiO$_{2}$ substrates and patterned into $2\,\mu$m wide wire devices (thickness in nm in parentheses, see Methods, SM1-2). The wires were connected through a circuit board to a pulse generator (\ref{fig:SkMotion-Setup}a) to inject current pulses of fixed width (20~ns), varying magnitude and polarity, $J=\pm(1.0-5.8)\times10^{11}$~A/m$^2$. The setup was mounted in an MFM with out-of-plane (OP) magnetic field ($\mu_{0}H$), enabling sequential \emph{in situ} pulse injection and imaging of the wire (\ref{fig:SkMotion-Setup}a-b). Pt/Co/MgO stacks are known to host N\'{e}el-textured skyrmions stabilized at RT by the interfacial Dzyaloshinskii-Moriya interaction \citep{Boulle2016,Juge2019}. Following an established nucleation recipe (see Methods, SM3), skyrmion configurations were stabilized in the wires over a substantial field range -- $\mu_{0}H\sim$75-165~mT -- as seen in MFM images (\ref{fig:SkMotion-Setup}f: inset). The range of skyrmion densities ($n_{{\rm S}}: 2-13\,\mu$m$^{-2}$) and sizes ($d_{\rm S}: 80-200$~nm) achieved here (details in SM3) are vastly different from previous works\citep{Jiang2016,Litzius2016,Woo2016,Woo2018,Hrabec2017,Legrand2017,Juge2019,Zeissler2020}, and provide the variance required to establish statistical significance for our key claims.

\fakesubsection{Detecting Skyrmion Motion}
 Current pulses of alternating polarity were applied sequentially to the wire (\ref{fig:SkMotion-Setup}b) and the motion of individual skyrmions was quantified by tracking their positions from MFM images acquired before and after each pulse (\ref{fig:SkMotion-Setup}c-e, details in SM4). \ref{fig:SkMotion-Setup}f summarizes the proportion of skyrmions,~$P_{\rm M}$, moving along the current direction ($J$) for various fields. We expect $P_{\rm M}$ to be positive for current-induced SOT displacement of left-handed N\'{e}el skyrmions stabilized in Pt/Co-based/MgO stacks\citep{Boulle2016,Woo2016,Woo2017}. For $J < 4\times10^{11}$~A/m$^2$,~$P_{\rm M}$ remains below 5\% -- characteristic of stochastic motion due to current-induced thermal fluctuations\citep{Emori2015,Legrand2017}. As $J$ is further increased, $P_{\rm M}$ increases exponentially -- reaching $\sim$90\% at $J\simeq 5.8\times10^{11}$~A/m$^2$ -- indicating a transition to driven skyrmion motion along the current direction. Finally, such deterministic motion can be further demarcated as creep or plastic flow, which is discussed further below.

\fakesection{\label{sec:SkHE}}

\begin{figure*}
\begin{centering}
\includegraphics[width=6.0in]{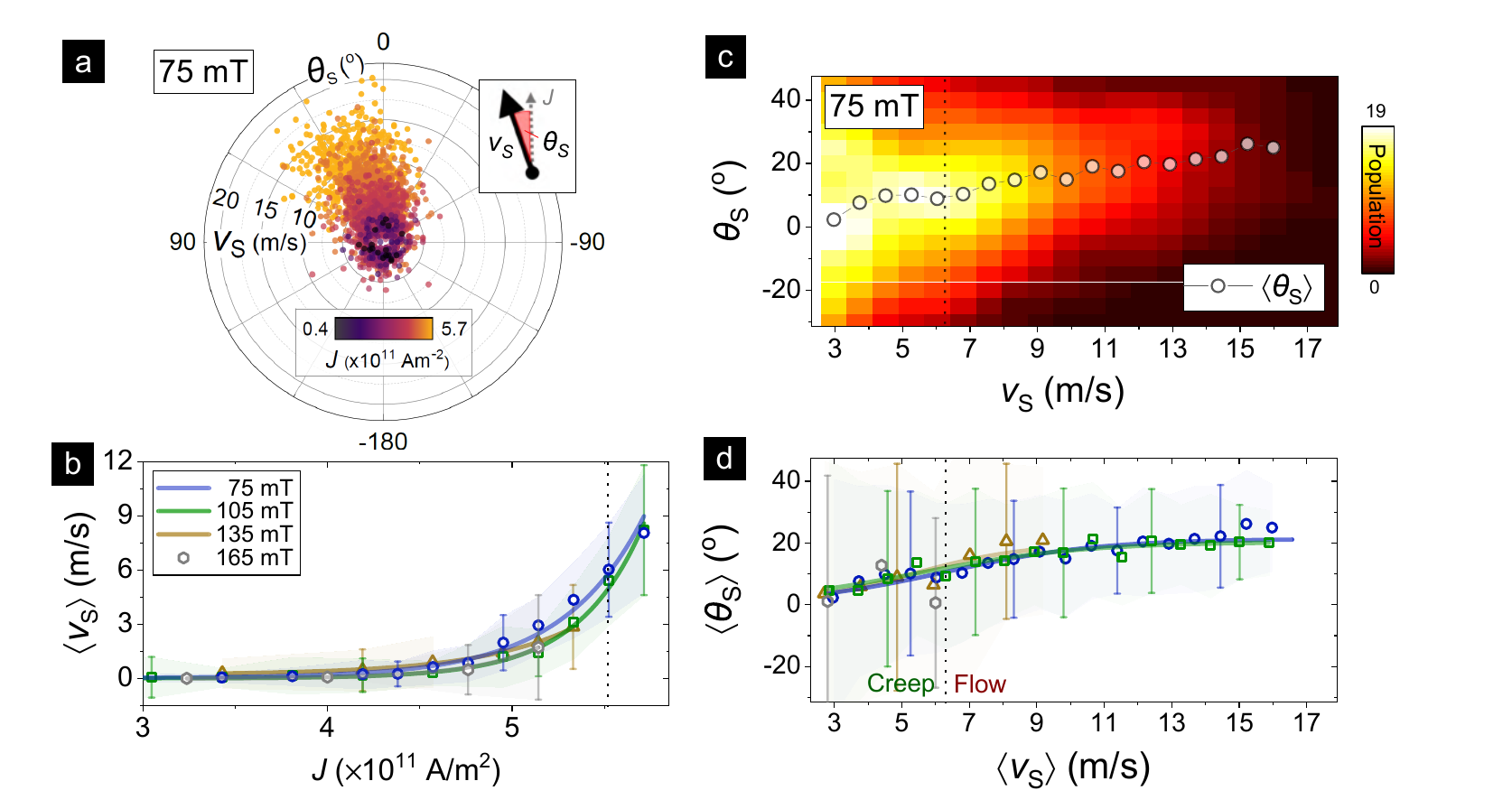}
\end{centering}
\caption[Current-Driven Skyrmion Dynamics]
{\textbf{Current-Driven Skyrmion Dynamics Overview.}
\textbf{(a)} Polar plot overview of the skyrmion motion statistics for $\mu_{0}H\simeq75$~mT across all currents $J$ (data for $J<0$ flipped by $180^\circ$), showing the spread of velocity ($v_{\rm S}$) and angular deflection ($\theta_{\rm S}$). The data is visibly biased towards the upper left quadrant. Inset shows a schematic defining $v_{\rm S}$ and $\theta_{\rm S}$ relative to the direction of $J$.
\textbf{(b)} Plot of the average skyrmion speed, $\langle v_{\rm S}\rangle$, against $J$ for various applied fields. Solid lines present exponential fits (see Methods), while shaded regions represent the standard deviation, also emphasized by error bars for selected points. The dotted line -- determined from (d) -- demarcates creep and plastic flow regimes.
\textbf{(c)} 2D histogram color plot of skyrmion deflection, $\theta_{\rm S}$ against $v_{\rm S}$ for $\mu_{0}H\simeq75$~mT across all currents. The data were binned by $v_{\rm S}$ for this plot. Solid markers show the average deflection, $\langle \theta_{\rm S}\rangle$, for each $v_{\rm S}$ bin.
\textbf{(d)} Plot of $\langle\theta_{\rm S}\rangle$ against $\langle v_{\rm S}\rangle$ -- determined as in (c) -- for various applied fields. Solid lines present sigmoidal fits (see Methods), while shaded regions represent the standard deviation, also emphasized by error bars for selected points. Creep and plastic flow regimes are demarcated at $\langle \theta_{\rm S} \rangle = 0.5\,\langle \theta_{\rm S} \rangle ^{\rm sat}$ using the fit.
\label{fig:Current-SkMotion}}\end{figure*}

\fakesubsection{Current-Velocity Relationship}
\textbf{Skyrmion Hall Effect.} The polar plot in \ref{fig:Current-SkMotion}a summarizes the distribution of skyrmion dynamics across applied currents at a representative field ($\mu_{0}H\simeq75$~mT). The variance in skyrmion velocity ($v_{\rm S}$) and angular deflection ($\theta_{\rm S}$) -- both defined with respect to $J$ (\ref{fig:Current-SkMotion}a: inset) -- is expected due to the granularity inherent to metallic multilayers \citep{Kim2017,Legrand2017}. Nevertheless, it is established that adequate statistical sampling and controls can meaningfully describe the skyrmion dynamics phenomenology \citep{Kim2017,Zeissler2020}. Correspondingly, \ref{fig:Current-SkMotion}b shows a plot of the averaged skyrmion velocity $\langle v_{\rm S}\rangle $ for each $J$. Notably, $\langle v_{\rm S}\rangle$ increases exponentially with $J$, reaching $\sim 8$~m/s for $J\simeq 5.8\times10^{11}$~A/m$^2$, with some skyrmions moving at over 24~m/s. The velocities observed here are of the same order of magnitude as Co-based skyrmionic counterparts subjected to similar $J$\citep{Woo2016,Hrabec2017,Juge2019}. Next, the threshold current for skyrmion motion increases with applied field, which can be ascribed to stronger pinning effects\citep{Kim2017,Legrand2017}. Importantly, however, the exponential trend of $\langle v_{\rm S}\rangle (J)$ is field-independent, indicating its insensitivity to variations in $n_{{\rm S}}$ for a given material system (c.f. \ref{fig:SkMotion-Setup}f, see SM3). However, with increasing field, the viable range of currents is substantially reduced (e.g. $J<5.2\times10^{11}$~A/m$^{2}$ for $\mu_{0}H\simeq165$~mT) due to the increased ease of skyrmion annihilation and stronger pinning\citep{Woo2016,Legrand2017}. Therefore, after demonstrating the consistency of skyrmion dynamics across fields (\ref{fig:SkMotion-Setup}-\ref{fig:Current-SkMotion}), the remainder of our work focuses -- without loss of generality -- on lower fields ($75-105$~mT) for statistically meaningful conclusions.

\fakesubsection{Skyrmion Hall Angle}
 Turning now to the angular deflection, skyrmion motion in \ref{fig:Current-SkMotion}a is noticeably skewed towards the upper left quadrant ($\theta_{\rm S} > 0$) -- increasingly so at higher velocities. The observed deflection is consistent with the expected SkHE, as skyrmions with topological charge $Q=+1$ should possess a positive intrinsic Hall angle ($\theta_{\rm H}> 0$)\citep{Nagaosa2013}. The 2D histogram plot in \ref{fig:Current-SkMotion}c shows the variation of $\langle\theta_{\rm S}\rangle$ with $\langle v_{\rm S}\rangle$ across currents. For lower values of $\langle v_{\rm S}\rangle$ e.g. $\lesssim 6$~m/s, $\langle\theta_{\rm S}\rangle$ is small and monotonically increasing while displaying a large variance. These attributes are characteristic of creep motion in a disordered background -- wherein scattering from the large fraction of pinned skyrmions results in a wide spread in $\langle\theta_{\rm S}\rangle $\citep{Reichhardt2016,Zeissler2020,Legrand2017}. As $\langle v_{\rm S}\rangle$ increases further, $\langle\theta_{\rm S}\rangle$ narrows in spread and grows in an S-curve fashion -- eventually saturating at $\theta_{\rm S}^{\rm{sat}}\sim 22^\circ$  for $\langle v_{\rm S}\rangle \gtrsim 12$~m/s. The $\langle\theta_{\rm S}\rangle$ saturation signals the onset of plastic flow, wherein skyrmions move concomitantly ($P_{\rm M} >$ 50\%) in a weak pinning background\citep{Reichhardt2016,Juge2019,Zeissler2020}.

\fakesubsection{Creep \& Flow Regimes}
 As shown in \ref{fig:Current-SkMotion}d, both the $\langle\theta_{\rm S}\rangle (v_{\rm S})$ profile and $\theta_{\rm S}^{\rm{sat}}$ magnitude are consistent over the entire field range ($75-165$~mT). This enables us to clearly demarcate the creep and plastic flow regimes in our work (\ref{fig:Current-SkMotion}d), which in turn establishes the current regime corresponding to plastic skyrmion flow (\ref{fig:Current-SkMotion}c: $J\geq 5.5\times 10^{11}$~A/m$^2$). Such a demarcation, while contrasting with some reports that solely exhibit either creep\citep{Litzius2016} or plastic flow\citep{Zeissler2020}, is consistent with other works\citep{Jiang2016,Juge2019,Woo2018}. Surprisingly, the saturated magnitude of $\theta_{\rm S}^{\rm{sat}}\sim 22^\circ$ is nearly $2-3$ times lower than the $40^\circ-70^\circ$ range found in most ferromagnetic multilayers\citep{Jiang2016,Litzius2016,Juge2019}, and is comparable to the $25^\circ-35^\circ$ values for ferrimagnetic systems\citep{Hirata2019,Woo2018}. Crucially, the established robustness with field enables us to examine extrinsic influences on the plastic flow of individual skyrmions in subsequent sections.

\fakesection{\label{sec:EdgeEffect}}
\begin{figure*}
\begin{centering}
\includegraphics[width=5in]{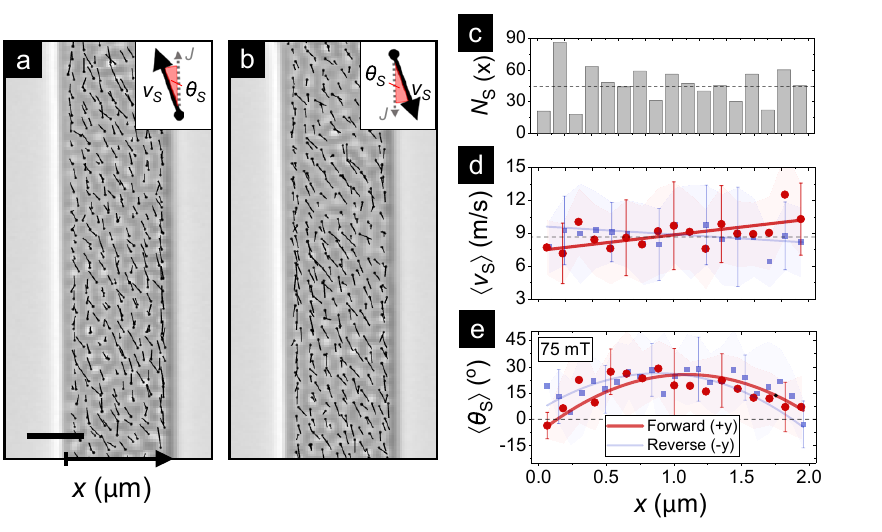}
\end{centering}
\caption[Edge Effect on Skyrmion Dynamics]{\textbf{Confinement Effects on Skyrmion Flow Dynamics}. 
\textbf{(a-b)} Representative MFM images (scalebar: $1\,\mu$m) at $\mu_0 H \simeq 75$~mT showing the forward ($J \parallel +\hat{y}$) (a) and reverse ($J \parallel -\hat{y}$) (b) plastic flow of skyrmions with varying distance $x$ from the left edge. Overlaid dots correspond to initial skyrmion positions and lines show the extent of motion due to the current pulse. Insets define $v_{\rm S}$ and $\theta_{\rm S}$ with respect to $J$ for both cases.
\textbf{(c)} Binned histogram distribution of skyrmions based on their $x$-positions before motion.
\textbf{(d-e)} Average velocity $\left\langle v_{{\rm S}}(x)\right\rangle$ (d) and angular deflection $\left\langle \theta_{{\rm S}}(x)\right\rangle $ (e) for skyrmions in each $x$-bin for forward ($J$\ensuremath{\parallel}+$\hat{y}$, red) and reverse ($J$\ensuremath{\parallel}-$\hat{y}$, blue) motion (details in SM7). Solid lines serve as guides-to-the-eye to indicate trends, while shaded regions represent the standard deviation, also emphasized by error bars for selected points. \label{fig:EdgeEffect}}
\end{figure*}

\fakesubsection{Edge Dependence of Skyrmion Density}
\textbf{Effect of Wire Edge.} Geometric confinement can strongly influence the stability and dynamics of skyrmions via magnetostatic and torque contributions\citep{Sampaio2013,Tomasello2014,Du2015a}. Previous theoretical and experimental works on single skyrmion in the creep regime have studied their interaction with a geometric boundary or ``edge'', and variously reported edge-induced skyrmion pinning\citep{Jiang2016,Lai2017}, annihilation\citep{Iwasaki2013,Lai2017,Bessarab2018},
expulsion\citep{Tomasello2014,Yoo2017,Bessarab2018} or repulsion\citep{Tomasello2014,Jiang2016,Iwasaki2013}. Here we examine the influence of confinement on the plastic flow of skyrmion arrays. Skyrmions in the plastic flow regime are binned by their individual distance $x$ from the left edge of the wire, with $0<x<2\,\mu$m (see SM7). The binned skyrmion number $N_{\rm S}(x)$ (\ref{fig:EdgeEffect}c) is approximately uniform across the $x$-bins, with no evidence for preferential existence at the centre or edge. Meanwhile \ref{fig:EdgeEffect}a-b show a marked evolution in the dynamics with varying $x$-positions, which is quantified in \ref{fig:EdgeEffect}d-e.

\fakesubsection{Edge Dependence of Velocity \& SkHE}
 \ref{fig:EdgeEffect}d shows that the binned velocity $\langle v_{\rm S}(x)\rangle$ remains consistently high ($\sim8-10$~m/s) across the wire width. A slight ($\sim$20\%) decrease of $\langle v_{\rm S}(x)\rangle$ is found at the left edge of the wire -- i.e. in the direction of skyrmion deflection. This suggests that skyrmion-edge interaction may have a measurable inelastic component -- consistent with a dissipative process\citep{Jiang2016,Iwasaki2013}. Meanwhile, the binned deflection $\langle\theta_{\rm S}(x)\rangle$ shows a parabolic evolution across the wire: increasing from $\sim -5^\circ$ at left to $\sim +25^\circ$ at centre (c.f. \ref{fig:Current-SkMotion}d, $\theta_{\rm S}^{\rm{sat}}\sim 22^\circ$) and then dropping to $\sim +10^\circ$ at the right. Crucially, we observe the same evolution of $\langle v_{\rm S}(x)\rangle$ and $\langle\theta_{\rm S}(x)\rangle$ when the motion is reversed (albeit with flipped sign: see \ref{fig:EdgeEffect}d-e) and at higher fields (105 mT, see SM6). Noting the consistency of edge effects across field variations of 30 mT, we rule out the role of current-induced Oersted fields in the observed trends.

\fakesubsection{Interplaying Effects on Position Dependent SkHE}
 Instead, the $\langle\theta_{\rm S}(x)\rangle$ variation may be interpreted within the interplay of the intrinsic Hall effect of individual skyrmions with extrinsic effects from the wire edge and neighbouring skyrmions. First, near the centre ($x\sim 1\,\mu$m), edge effects are negligible and neighbouring skyrmion effects are compensated on both sides. Therefore, the ensuing $\langle\theta_{\rm S}\rangle$ is comparable to the saturation value (see Results: Skyrmion Hall Effect). Next, at the left edge $(x\sim 0\,\mu$m), the Magnus force is overcome by edge repulsion. The latter pushes the skyrmion back into the wire -- resulting in a negative $\langle\theta_{\rm S}\rangle$. Finally, the gradual reduction of $\langle\theta_{\rm S}\rangle$ for ($1<x<2\,\mu$m) cannot be explained by edge effects. Instead, it suggests that the gradual reduction of skyrmions on the right -- whose transverse motion may repel the skyrmion of interest -- may influence the magnitude of $\langle\theta_{\rm S}\rangle$ observed in our work. Finally, the robustness of these results to field and current direction underscores the deterministic role of extrinsic factors in shaping skyrmion dynamics in our wires.

\fakesection{\label{sec:Irrev-SampleDep}}

\begin{figure*}[t]
\begin{centering}
\includegraphics[width=6.2in]{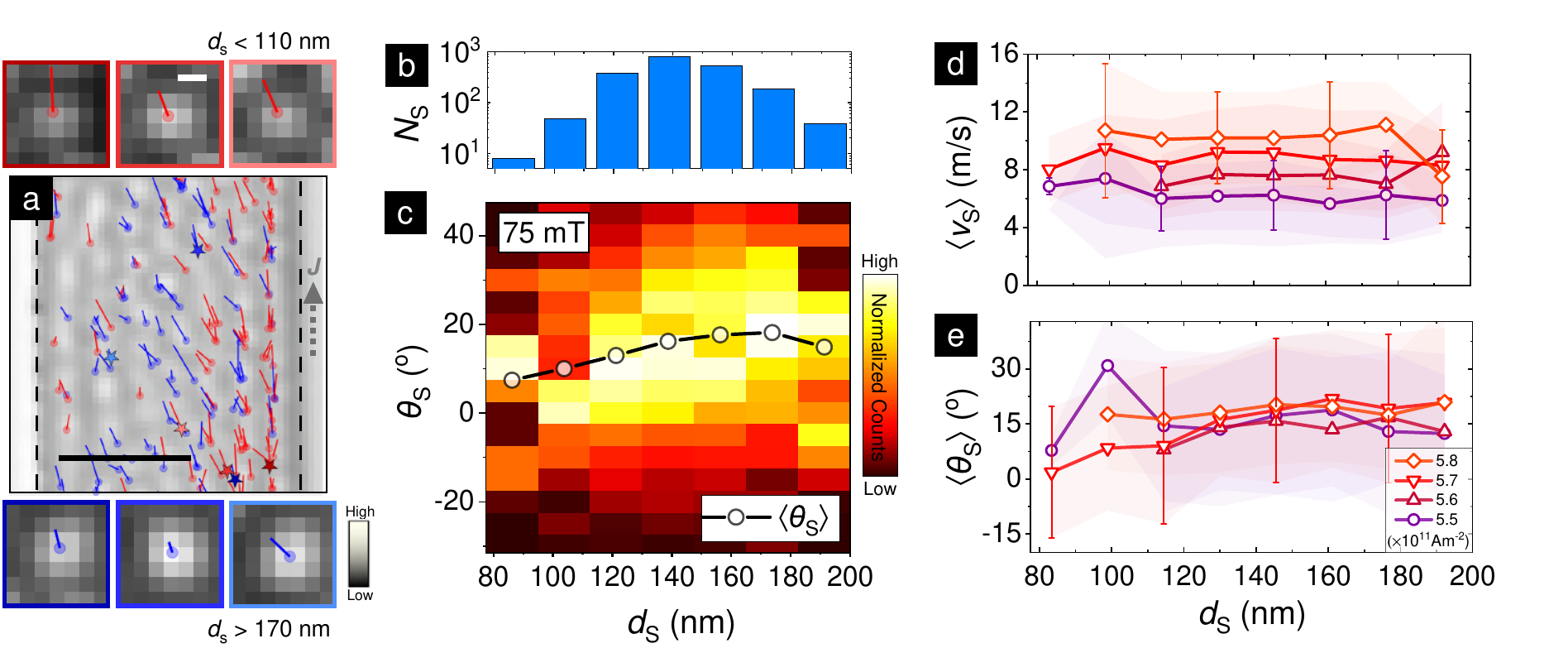}
\end{centering}
\caption[Skyrmion Size Effect Experiments]{\textbf{Skyrmion Size Effect on Flow Dynamics.} 
\textbf{(a)} Representative cropped MFM image (scalebar: $1\,\mu$m) at $\mu_{0}H\simeq 75$~mT showing the evolution in the plastic flow of skyrmions with varying size $d_{\rm S}$. The smallest 25\% (red) and largest 25\% (blue) of $d_{\rm S}$ are highlighted. Overlaid dots correspond to initial skyrmion positions and lines show the extent of motion. Zoomed insets at top and bottom (scalebar: 100~nm) show selected small (top, $<110$~nm) and large (bottom, $>170$~nm) skyrmions for comparison. 
\textbf{(b)} Binned histogram distribution of skyrmions, $N_{\rm S}$, based on their size, $d_{\rm S}$ -- which varies over $80-200$~nm (details in SM7). 
\textbf{(c)} 2D histogram color plot of skyrmion deflection, $\theta_{\rm S}$ against $d_{\rm S}$ for $\mu_{0}H\simeq75$~mT across all currents in the plastic flow regime. The data were binned by $d_{\rm S}$ and normalized in each bin for this plot. Solid markers show the average deflection, $\langle\theta_{\rm S}\rangle$, for each $d_{\rm S}$ bin.
\textbf{(d-e)} Average velocity $\langle v_{\rm S}\rangle$ (d) and angular deflection $\langle\theta_{\rm S}\rangle$ (e) for skyrmions in each $d_{\rm S}$-bin for $J$ over $(5.5-5.8)\times10^{11}$~A/m$^2$. Shaded regions represent the standard deviation, also emphasized by error bars for selected points. 
\label{fig:SkSizeEffect-Expts}}\end{figure*}

\fakesubsection{Size Dependence of Velocity}
\textbf{Effect of Skyrmion Size.} The size of a skyrmion determines its coupling to current-induced spin torques, and is therefore expected to influence its dynamics\citep{Thiele1973,Locatelli2013}. In \ref{fig:SkSizeEffect-Expts}, we examine the skyrmion size dependence seen in our experiments in the plastic flow regime for $\mu_{0}H \simeq 75$~mT and $J$ over $(5.5-5.8)\times 10^{11}$~A/m$^2$. The skyrmions are binned by their $d_{\rm S}$ in MFM images (see SM7), which show a substantial spread over 80--200~nm (\ref{fig:SkSizeEffect-Expts}a insets, \ref{fig:SkSizeEffect-Expts}b). First, \ref{fig:SkSizeEffect-Expts}d shows that $\langle v_{\rm S}\rangle$ -- while increasing as expected with $J$ -- is constant to $\sim$10\% across $d_{\rm S}$ for fixed $J$. Such insensitivity of $\langle v_{\rm S}\rangle$ to $d_{\rm S}$, predicted for $d_{\rm S} > 100$~nm and recently reported for $\langle d_{\rm S}\rangle\sim 400$~nm skyrmion bubbles\citep{Zeissler2020}, may be ascribed to the onset of finite size effects within the skyrmion spin structure\citep{Legrand2017}.

\fakesubsection{Size Dependence of SkHE}
 Meanwhile, the variation of $\langle\theta_{\rm S}\rangle$ with $d_{\rm S}$ across currents is shown in the 2D histogram plot in \ref{fig:SkSizeEffect-Expts}c. $\langle\theta_{\rm S}\rangle$ increases discernibly with $d_{\rm S}$ across all $J$-values in the plastic flow regime (\ref{fig:SkSizeEffect-Expts}e) -- from $\sim 5-10^\circ$ for $d_{\rm S}\lesssim 100$~nm to $\sim 20^\circ$ for $d_{\rm S} \sim 200$~nm. While the limited current range for plastic flow ($5.5-5.8\times 10^{11}$~A/m$^2$) precludes meaningful $J$-dependent trend, the weakly increasing trend of $\langle\theta_{\rm S}\rangle$ with $d_{\rm S}$ noted here is robust to binning (SM7). Moreover, similar results are observed for 105~mT (SM6, 7), and another wire device as well (SM6). Our results contrast strongly with the $1/d_{\rm S}$ dependence of $\theta_{\rm H}$ expected theoretically from the Thiele model for rigid skyrmions\citep{Thiele1973,Knoester2014}, and with experimental reports of such a $\langle \theta_{\rm S}\rangle$ trend in the creep regime\citep{Litzius2016}. Meanwhile, a recent work has reported $d_{\rm S}$-independent $\langle\theta_{\rm S}\rangle$ in the plastic flow regime\citep{Zeissler2020}. In light of contrasting theoretical and experimental reports, the size dependence of SkHE has increasingly assumed importance. Our simulations and theoretical work attempt to identify plausible origins of these discrepancies.

\begin{figure}
\begin{centering}
\includegraphics[width=2.6in]{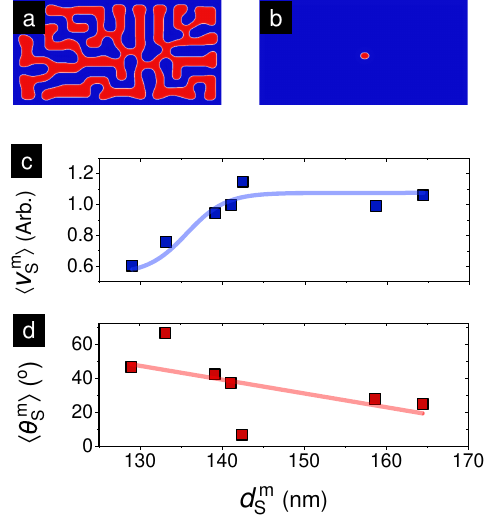}
\end{centering}
\caption[Micromagnetic Simulations of Skyrmion Size Effects]{\textbf{Micromagnetic Simulations of Skyrmion Size Effects.} \textbf{(a-b)} Representative micromagnetic magnetization of a $2\times 4\,\mu$m$^2$ wire showing a single skyrmion formed upon relaxation at $\mu_0 H\simeq 85$~mT. The grain-free simulations were performed using stack parameters consistent with experiments (details in Methods).
\textbf{(c-d)} Evolution of micromagnetic simulated average velocity $\langle v_{\rm S}^{\rm m}\rangle$ (c) and average angular deflection $\langle \theta_{\rm S}^{\rm m}\rangle$ (d) with $d_{\rm S}$ -- extracted from a series of such simulations at a representative $J=9.5\times10^{11}$~A/m$^2$. Solid lines represent sigmoidal (c) and linear (d) fits, respectively.
\label{fig:SizeEffectSims-uMag}}\end{figure}

\begin{figure}
\begin{centering}
\includegraphics[width=2.4in]{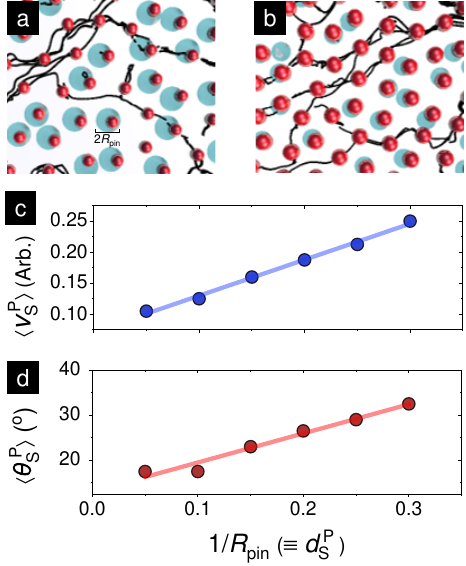}
\end{centering}
\caption[Particle Model Simulations of Skyrmion Size Effects]{\textbf{Particle Model Simulations of Skyrmion Size Effects.}\textbf{(a-b)} Schematics of particle model simulations (details in Methods) which examine the dynamics of point-like skyrmion particles (foreground, red) in the presence of pinning sites with radius $R_{\rm pin}$ (background, blue). Black lines show representative trajectories of skyrmions. Decreasing $R_{\rm pin}$ within the model (left to right) emulates the effect of increasing $d_{\rm S}$.
\textbf{(c-d)} Evolution of particle model simulated average velocity $\langle v_{\rm S}^{\rm p}\rangle$ (c) and average angular deflection $\langle \theta_{\rm S}^{\rm p}\rangle$ (d) with $1/R_{\rm pin} \equiv d_{\rm S}$ extracted from a series of particle model simulations with fixed drive. 
\label{fig:SizeEffectSims-part}}\end{figure}

\fakesubsection{Micromagnetic Simulations}
\textbf{Micromagnetics.} Micromagnetic simulations were performed for a grain-free environment using stack parameters consistent with experiments to study the size dependence of the intrinsic SkHE (see Methods, SM8)\citep{Vansteenkiste2014}. Skyrmions were stabilized in a $2\times 4 \: \mu$m wire geometry with varying $\mu_0 H$ (84--93~mT, see e.g. \ref{fig:SizeEffectSims-uMag}a-b), which resulted in a 20\% variation in simulated $d_{\rm S}$ over 129--164~nm. Analysis of the current-induced skyrmion motion, shown in \ref{fig:SizeEffectSims-uMag}c-d for $J=9.5\times10^{11}$~A/m$^2$, reveals that with increasing $d_{\rm S}^{\rm m}$, the simulated $\langle v_{\rm S}^{\rm m}\rangle$ initially increases and eventually saturates for $d_{\rm S}\gtrsim 150$~nm. Meanwhile $\langle\theta_{\rm S}^{\rm m}\rangle$ decreases from $40^\circ$ to $20^\circ$. Both these trends are consistent with the expected rigid skyrmion behaviour and with recent simulations in the creep regime\citep{Litzius2016,Legrand2017}. However, neither our grain-free micromagnetic simulations nor those incorporating inhomogeneity\citep{Juge2019,Kim2017,Legrand2017} -- can explain the $d_{\rm S}$-dependent $\langle\theta_{\rm S}^m\rangle$ trends in the plastic flow regime seen in \ref{fig:SkSizeEffect-Expts}, or in other recent works\citep{Zeissler2020}. Therefore, we turn to the particle model -- an established technique for elucidating the dynamics of skyrmion arrays in a disordered background (see Methods)\citep{Reichhardt2016,Reichhardt2015}.

\fakesubsection{Particle Model Setup}
\textbf{Particle Model.} As shown in \ref{fig:SizeEffectSims-part}a-b, skyrmions are represented as an array of point particles, while the disorder is modelled using pinning sites of radius, $R_{\rm pin}$ (details in Methods). Within this model, the effectiveness of pinning scales with the area coverage, given by $(d_{\rm S})^2 / (R_{\rm pin})^2$, where $d_{\rm S}$ is the skyrmion size. Therefore, the impact of changing skyrmion size can be modelled either by increasing/decreasing $d_{\rm S}$ or by increasing/decreasing  $1/R_{\rm pin}$. Here, we increase $R_{\rm pin}$ to emulate the effect of reducing  $d_{\rm S}$ (\ref{fig:SizeEffectSims-part}a), wherein smaller skyrmions would experience a larger interaction landscape with a given pinning site. The $d_{\rm S}^{\rm p}$ in \ref{fig:SizeEffectSims-part}c-d is described as a dimensionless number based on the ratio of the $d_{\rm S}$ to $R_{\rm pin}$. The simulations were seeded with 1,500 particles and 1,200 pinning sites, and the system was subjected to a fixed drive corresponding to plastic flow under the modified Thiele equation (see Methods). The intrinsic, or pin-free skyrmion Hall angle, $\theta_{\rm H}$, was set to $37^\circ$, while $R_{\rm pin}$ was varied over a wide range to simulate the role of pinning effects in the observed size dependence of skyrmion dynamics.

\fakesubsection{Particle Model Results}
 On one hand, \ref{fig:SizeEffectSims-part}d shows that the particle model angular deflection, $\langle\theta_{\rm S}^{\rm p}\rangle$ decreases monotonically with reducing $d_{\rm S}$ (simulated by increasing $R_{\rm pin}$) to values substantially below the intrinsic Hall angle ($\theta_{\rm H} = 37^\circ$). Additional simulations for $\theta_{\rm H} = 45^\circ$ confirm this trend, which is qualitatively consistent with experiments. The simulations suggest that while smaller skyrmions may in fact have a higher intrinsic $\theta_{\rm H}$, as per the Thiele model\citep{Thiele1973,Knoester2014}, extrinsic interactions may considerably reshape the effective $\langle \theta_{\rm S}^{\rm p}\rangle$. In particular, the stronger influence of pinning on smaller skyrmions may impede their transverse deflection, reducing $\langle\theta_{\rm S}^{\rm p}\rangle$ to well below the intrinsic value. Meanwhile the simulation results also suggest the $d_{\rm S}$-independent SkHE reported in another recent work\citep{Zeissler2020} may result from the dominance of pinning effects over the intrinsic Hall effect. In this case, the effective pinning could be acting at scales much larger than $d_{\rm S}$, and could therefore result in the diminished contribution of variations in $d_{\rm S}$. On the other hand, \ref{fig:SizeEffectSims-part}c shows that the particle model velocity, $\langle v_{\rm S}^{\rm p}\rangle$ increases linearly with $1/R_{\rm pin} \equiv d_{\rm S}$, likely because increased pinning would slow down point-like skyrmions. In comparison, the $d_{\rm S}$-independence of $v_{\rm S}$ in our experiments and previous work\citep{Zeissler2020} may be attributed, from micromagnetic simulations, to the onset of bubble-like internal structure of skyrmions\citep{Legrand2017}. Such textural size effects, whose internal modes may, for smaller $d_{\rm S}$, also reduce the effective magnitude of adiabatic and non-adiabatic damping, are beyond the scope of these particle model simulations. Future theoretical works may incorporate these effects and consider the skyrmion interactions arising from the vastly higher magnitude of skyrmion densities in our experiments. Nevertheless, these insights provide an important stepping stone for delineating intrinsic and extrinsic contributions to skyrmion dynamics.

\section*{Discussion\label{sec:Outlook}}

\fakesubsection{Results Summary}
In summary, we have presented a systematic study of the ensemble dynamics of skyrmions forming dense array configurations in multilayer wire devices. With increasing current, we observe distinct transitions in skyrmion dynamics -- from stochastic to deterministic creep, accompanied by SkHE onset, and then to plastic flow -- wherein SkHE saturates at $\sim 22^\circ$. Within the plastic flow regime, we find that the skyrmion velocity is surprisingly robust to edge effects and skyrmion size, while the SkHE varies considerably in response to both. Notably, the weak increase of SkHE with skyrmion size contradicts existing predictions for isolated, rigid skyrmions. Instead, we conclude that the intrinsic Hall effect of skyrmions is strongly reshaped by extrinsic contributions in our wires -- including geometric confinement, disorder-induced pinning effects and skyrmion-skyrmion interactions.

\fakesubsection{Impact}
 The insights from our work dispel several prevailing notions on skyrmion dynamics in multilayers, and pave imminent materials and device directions for high throughput racetrack devices. First, the relatively small magnitude of SkHE in our case (saturated, $\sim 22^\circ$) is comparable to ferrimagnetic multilayers\citep{Woo2018}. This suggests that materials design efforts for optimizing SkHE should continue exploring conventional ferromagnets in addition to newer compensated systems. Second, we find the skyrmion-edge interaction to be weakly inelastic: unlike several predictions\citep{Yoo2017,Lai2017,Bessarab2018}, the edge does not annihilate or pin skyrmions. In fact, such skyrmion-edge interactions may be exploited to channel skyrmion motion within suitably designed device geometries\citep{Luo2018}. Third, size effects on skyrmion dynamics are reshaped by material granularity and skyrmion-skyrmion interactions. Contrary to defect-free theoretical predictions, in our case smaller skyrmions move equally fast, and importantly with reduced SkHE. While emphasizing accurate inclusion of these effects in future theoretical works, we postulate that they may serve as tuning parameters to achieve bespoke size-dependent skyrmion dynamics in racetrack devices\citep{Parkin2015,Zhao2012}. Finally, we note that the striking individuality of our skyrmions within dense, disordered arrays bodes well for their use as stochastic spiking neurons in synaptic computing applications\citep{Pinna2018,Li2017}.
\section*{Methods\label{sec:Methods}}

\begin{small}
\fakesubsection{Film Deposition}
\textbf{Film deposition.} Multilayer stacks of\\ Ta(3)/[Pt(3)/\textbf{Co(1.2)}/MgO(1.5)]$_{15}$/Ta(4) (nominal layer thicknesses in nm in parentheses) were deposited on pre-cleaned thermally oxidized 200~mm Si wafers by ultrahigh vacuum magnetron sputtering at RT using the Singulus Timaris\texttrademark{} system. Magnetization measurements were performed on the film using the MicroSense\texttrademark{} Model EZ11 vibrating sample magnetometer. An effective OP anisotropy, $K_{\rm eff}$ of 0.17~MJ/m$^3$ and saturation magnetization, $M_{\rm s}$ of 1.1~MA/m were determined from the $M(H)$ data. The DMI, $D$ and exchange stiffness, $A$ of the film were determined to be 1.6 mJ/m$^2$ and 24~pJ/m, respectively, using established techniques\citep{MoreauLuchaire2016,Woo2016,Soumyanarayanan2017}. These values are in line with published results on Pt/Co(0.9)/MgO stacks\citep{Boulle2016,Juge2019}.

\fakesubsection{Device Fabrication}
\textbf{Device fabrication.} A 300~nm thick negative resist, Ma-N 2403, was spin-coated on the multilayer film. Wires of dimensions $2\times 10$~$\mu$m were exposed using the Elionix\texttrademark{} electron beam lithography tool. The patterns were transferred onto the multilayer film using an Oxford CAIBE\texttrademark{} ion beam etching system, and residual resist was lifted off in an ultrasonic bath. Top electrodes were subsequently patterned using an EVG\texttrademark{} optical mask aligner, followed by the deposition of the electrode stack Ta(5)/Au(100)/Ru(20) using the Chiron\texttrademark{} UHV magnetron sputtering system.

\fakesubsection{MFM and Pulsing Setup}
\textbf{MFM and pulsing setup.} MFM imaging was performed using a Veeco Dimension\texttrademark{} 3100 scanning probe microscope with Co-alloy coated SSS-MFMR\texttrademark{} tips. The sharp tip profile (diameter $\sim 30$~nm), ultra-low moment ($\sim 80$~emu/cm$^3$), and lift heights of 20--30~nm used during scanning provided high-resolution MFM images while introducing minimal stray field perturbations. Our earlier works have established MFM as a reliable tool for imaging sub-100 nm skyrmions in multilayer films\citep{Soumyanarayanan2017,Ho2019}. \textit{In situ} electrical pulsing and imaging were carried out in our custom-designed platform consisting of a Tektronix\texttrademark{} AFG3252 pulse generator, SRS\texttrademark{} SIM954 inverting amplifier, Tektronix\texttrademark{} TDS 7404 oscilloscope and the microscope. The device under test was wire-bonded to the chip carrier and subsequently mounted onto the MFM setup with varying \textit{in situ} OP magnetic fields of 75--165~mT, following \textit{ex situ} negative OP saturation. The device was impedance matched and found to be $\sim 50~\Omega$. Ambipolar pulses of amplitude 1--3.7~V were injected, corresponding to current densities of $(1.0-5.8)~\times 10^{11}$~A/m$^2$ assuming a total metallic layer thickness of 63~nm. Short pulse duration of 20~ns was used to limit Joule heating effects on skyrmion nucleation, deletion and motion. The resistance and zero field configuration of the devices were verified prior to and after pulsing, affirming that the pristine device form was preserved over the course of pulse injection experiments.

\fakesubsection{Pulsing Experiments}
\textbf{Pulsing experiments.} Magnetic configurations consisting solely of skyrmions were stabilized over these fields by injecting bipolar current pulses of magnitude $J < 5.5\times 10^{11}$A/m$^2$ (details in SM3). The procedure was repeated until all stripes were broken up into skyrmions, and no further skyrmions could be created. Subsequently, current-driven skyrmion dynamics experiments were performed with $J$ ranging over $(1.0-5.8)~\times 10^{11}$~A/m$^2$. Skyrmion motion was analysed by identifying and tracking the skyrmion positions on the MFM image after each pulse. The $\langle v_{\rm S} \rangle$ and $\langle \theta_{\rm S}\rangle$ were extracted by calculating their average displacement over an effective pulse duration of 20~ns, taking into account the rise and fall times. The $\langle v_{{\rm S}}\rangle$ vs $J$ (\ref{fig:Current-SkMotion}b) and $\langle \theta_{\rm S}\rangle$ vs $\langle v_{\rm S}\rangle$ (\ref{fig:Current-SkMotion}d) plots are fitted using the exponential (\ref{eq:Exponential fit-1}) and sigmoidal (\ref{eq:sigmoidal fit}) functions respectively, defined as
\begin{equation}
y=a+b\exp\left(\frac{x-c}{d}\right)\label{eq:Exponential fit-1}
\end{equation}
\begin{equation}
y=\frac{a}{1+\exp\left(-\frac{x+b}{c}\right)}+d\label{eq:sigmoidal fit}
\end{equation}
where $a$, $b$, $c$ and $d$ are constants.

\fakesubsection{Skyrmion Dynamics Analysis}
\textbf{Skyrmion dynamics analysis.} To ensure that the devices imaged after pulsing were at identical positions for reliable tracking and analysis of skyrmion motion, an image registration protocol established using the image processing toolbox in MATLAB$^\circledR$ was implemented. The MFM images obtained from the consecutive pulses were aligned by performing a 2D geometric transformation consisting of translation, rotation and shear relative to a reference image (described in SM4). Following image alignment, all skyrmions were identified and each was tagged with a unique marker. Next, each skyrmion was traced to its new position after pulsing from its original position by systematically tracking around its nearest position starting from top-down (or bottom-up, depending on the pulse direction). The rigorous tracking protocol accounts for nearly all skyrmion motion, as the local skyrmion number is typically unchanged through the pulsing experiments (see SM4).

\fakesubsection{Micromagnetic Simulations}
\textbf{Micromagnetic simulations.} Micromagnetic simulations were performed using the MuMax3 software\citep{Vansteenkiste2014} on a rectangular area of $2\times 4$~$\mu$m$^2$ to mimic the experimental wire structure. In view of computational constraints, the 15 repeat stack was simulated with an effective medium model\citep{Woo2016}, wherein each repeat was represented by one effective FM layer. The magnetic parameters used in the effective medium model were rescaled from the experimentally measured magnetic parameters, where $A$ = 5.05~pJ/m, $M_{\rm s}$ = 0.23~ MA/m, $K_{\rm eff}$ = 0.070~MJ/m$^3$ and $D$ = 0.37~mJ/m$^2$\citep{Woo2016}. An experimentally determined Gilbert damping parameter of $\alpha$ = 0.05 was used in this simulation (details in SM1). To simulate the skyrmion dynamics, the spin-orbit torque on the Co layer was modelled as an anti-damping-like torque from the adjacent Pt layer with an effective spin-Hall angle of 0.1. For simplicity, field-like torque originating from Pt was not considered and simulations were carried out at zero-temperature. Additionally, the role of thermal heating was neglected in line with experimental observation of negligible current induced heating effects (see SM3).

\fakesubsection{Particle Model Simulations}
\textbf{Particle model simulations.} Skyrmion dynamics within the particle model was simulated using a modified Thiele equation\citep{Reichhardt2016,Reichhardt2015}
\begin{equation}
\alpha_{\rm d}v_{i}+\alpha_{\rm m}\hat{Z}\times v_{i}=F_{i}^{\rm ss}+F_{i}^{\rm sp}+F^{\rm D}\label{eq:Particle-Thiele}
\end{equation}
Here, $\alpha_{\rm d}$ is the damping and $\alpha_{\rm m}$ is the Magnus force. The term $F_{i}^{\rm ss}$ is the skyrmion repulsive interaction, and $F_{i}^{\rm sp}$ is the skyrmion-pinning interaction. The pinning was modelled as arising from localized sites of radius $R_{\rm pin}$ with a finite range harmonic attractive potential which gives a maximum pinning force strength of $F_{\rm pin}$. The term $F^{\rm D}$ represents a dc drive on the skyrmions applied in the $x$-direction. The skyrmion velocity parallel and perpendicular to the drive is $v_{{\rm S},\parallel}^{\rm P}$ and $v_{{\rm S},\perp}^{{\rm P}}$ respectively, while the measured skyrmion Hall angle is $\theta_{S}^{\rm P}=\tan^{-1}(v_{{\rm S},\perp}^{\rm P}/v_{{\rm S},\parallel}^{\rm P})$. In the absence of disorder, the skyrmion Hall angle is $\tan^{-1}(\alpha_{\rm m}/\alpha_{\rm d})$. The simulations presented here were performed with the ratio of the number of pinning sites to the number of skyrmions being 0.6. To mimic the effect of changing the skyrmion diameter, the pinning force was held constant, and the pinning radius $R_{\rm pin}$ was varied. In this case, a large pinning site would correspond to a smaller skyrmion. The effective skyrmion Hall angle and the skyrmion velocity were then measured from the simulations. The values of $F_{\rm pin}=1.0$, $\alpha_{\rm d}=1.34$, and $\alpha_{\rm m}=1.0$ were used for simulations, giving an intrinsic Hall angle of 37$^\circ$.

\end{small} 

\noindent
\begin{center}{\rule[0.5ex]{0.4\columnwidth}{0.5pt}}\end{center}

\begin{small}
\noindent \textsf{\textbf{Acknowledgments.}} 
We thank Qi Jia Yap, Sze Ter Lim, and Franck Ernult for valuable experimental inputs, and Soong-Geun Je, Mi-Young Im, and Xichao Zhang for insightful discussions. We acknowledge the support of the National Supercomputing Centre (NSCC) for computational resources. This work was supported by the SpOT-LITE programme (Grant Nos. A1818g0042, A18A6b0057), funded by Singapore's RIE2020 initiatives, and by the Pharos Skyrmion programme (Grant No. 1527400026) funded by A{*}STAR, Singapore. Further, we gratefully acknowledge the support of the U.S. Department of Energy through the LANL/LDRD program for this work. This work was also supported by the US Department of Energy through the Los Alamos National Laboratory. Los Alamos National Laboratory is operated by Triad National Security, LLC, for the National Nuclear Security Administration of the U. S. Department of Energy (Contract No. 892333218NCA000001). 

\noindent \textsf{\textbf{Data Availability.}} The data generated during and/or analysed during the current study are available from the corresponding author(s) on reasonable request.

\vspace{1ex}
\noindent \textsf{\textbf{Author Contributions.}} A.K.C.T., P.H., and A.S. designed and initiated the research. A.K.C.T. and J.L. designed the \emph{in situ} device imaging setup. P.H. and L.S.H. fabricated the devices. H.K.T. conducted magnetometry measurements, while J.L. carried out the damping measurements. A.K.C.T. carried out the MFM experiments and analyzed the imaging data. J.L. performed the micromagnetic simulations and analyzed the data with A.K.C.T and P.H. C.J.O.R and C.R. performed and analyzed the particle model simulations. P. H. and A.S. coordinated and supervised the project. All authors discussed the results and provided inputs to the manuscript.

\noindent \textsf{\textbf{Competing interests.}} The authors declare no competing interests

\end{small}
\noindent \bibliographystyle{naturemag}
\bibliography{SkWires}

\begin{thebibliography}{10}
\expandafter\ifx\csname url\endcsname\relax
  \def\url#1{\texttt{#1}}\fi
\expandafter\ifx\csname urlprefix\endcsname\relax\def\urlprefix{URL }\fi
\providecommand{\bibinfo}[2]{#2}
\providecommand{\eprint}[2][]{\url{#2}}

\bibitem{Nagaosa2013}
\bibinfo{author}{Nagaosa, N.} \& \bibinfo{author}{Tokura, Y.}
\newblock \bibinfo{title}{{Topological properties and dynamics of magnetic
  skyrmions}}.
\newblock \emph{\bibinfo{journal}{Nature Nanotechnology}}
  \textbf{\bibinfo{volume}{8}}, \bibinfo{pages}{899--911}
  (\bibinfo{year}{2013}).

\bibitem{Wiesendanger2016}
\bibinfo{author}{Wiesendanger, R.}
\newblock \bibinfo{title}{{Nanoscale magnetic skyrmions in metallic films and
  multilayers: a new twist for spintronics}}.
\newblock \emph{\bibinfo{journal}{Nature Reviews Materials}}
  \textbf{\bibinfo{volume}{1}}, \bibinfo{pages}{16044} (\bibinfo{year}{2016}).

\bibitem{Boulle2016}
\bibinfo{author}{Boulle, O.} \emph{et~al.}
\newblock \bibinfo{title}{{Room-temperature chiral magnetic skyrmions in
  ultrathin magnetic nanostructures}}.
\newblock \emph{\bibinfo{journal}{Nature Nanotechnology}}
  \textbf{\bibinfo{volume}{11}}, \bibinfo{pages}{449--454}
  (\bibinfo{year}{2016}).

\bibitem{MoreauLuchaire2016}
\bibinfo{author}{Moreau-Luchaire, C.} \emph{et~al.}
\newblock \bibinfo{title}{{Additive interfacial chiral interaction in
  multilayers for stabilization of small individual skyrmions at room
  temperature}}.
\newblock \emph{\bibinfo{journal}{Nature Nanotechnology}}
  \textbf{\bibinfo{volume}{11}}, \bibinfo{pages}{444--448}
  (\bibinfo{year}{2016}).

\bibitem{Woo2016}
\bibinfo{author}{Woo, S.} \emph{et~al.}
\newblock \bibinfo{title}{{Observation of room-temperature magnetic skyrmions
  and their current-driven dynamics in ultrathin metallic ferromagnets}}.
\newblock \emph{\bibinfo{journal}{Nature Materials}}
  \textbf{\bibinfo{volume}{15}}, \bibinfo{pages}{501--506}
  (\bibinfo{year}{2016}).

\bibitem{Soumyanarayanan2017}
\bibinfo{author}{Soumyanarayanan, A.} \emph{et~al.}
\newblock \bibinfo{title}{{Tunable Room Temperature Magnetic Skyrmions in
  Ir/Fe/Co/Pt Multilayers}}.
\newblock \emph{\bibinfo{journal}{Nature Materials}}
  \textbf{\bibinfo{volume}{16}}, \bibinfo{pages}{898--904}
  (\bibinfo{year}{2017}).

\bibitem{Miron2010a}
\bibinfo{author}{Miron, I.~M.} \emph{et~al.}
\newblock \bibinfo{title}{{Current-driven spin torque induced by the Rashba
  effect in a ferromagnetic metal layer}}.
\newblock \emph{\bibinfo{journal}{Nature Materials}}
  \textbf{\bibinfo{volume}{9}}, \bibinfo{pages}{230--4} (\bibinfo{year}{2010}).

\bibitem{Miron2011}
\bibinfo{author}{Miron, I.~M.} \emph{et~al.}
\newblock \bibinfo{title}{{Perpendicular switching of a single ferromagnetic
  layer induced by in-plane current injection}}.
\newblock \emph{\bibinfo{journal}{Nature}} \textbf{\bibinfo{volume}{476}},
  \bibinfo{pages}{189--193} (\bibinfo{year}{2011}).

\bibitem{Sampaio2013}
\bibinfo{author}{Sampaio, J.}, \bibinfo{author}{Cros, V.},
  \bibinfo{author}{Rohart, S.}, \bibinfo{author}{Thiaville, A.} \&
  \bibinfo{author}{Fert, A.}
\newblock \bibinfo{title}{{Nucleation, stability and current-induced motion of
  isolated magnetic skyrmions in nanostructures}}.
\newblock \emph{\bibinfo{journal}{Nature Nanotechnology}}
  \textbf{\bibinfo{volume}{8}}, \bibinfo{pages}{839--844}
  (\bibinfo{year}{2013}).

\bibitem{Jiang2015}
\bibinfo{author}{Jiang, W.} \emph{et~al.}
\newblock \bibinfo{title}{{Blowing magnetic skyrmion bubbles}}.
\newblock \emph{\bibinfo{journal}{Science}} \textbf{\bibinfo{volume}{349}},
  \bibinfo{pages}{283--286} (\bibinfo{year}{2015}).

\bibitem{Parkin2008}
\bibinfo{author}{Parkin, S. S.~P.}, \bibinfo{author}{Hayashi, M.} \&
  \bibinfo{author}{Thomas, L.}
\newblock \bibinfo{title}{{Magnetic domain-wall racetrack memory}}.
\newblock \emph{\bibinfo{journal}{Science}} \textbf{\bibinfo{volume}{320}},
  \bibinfo{pages}{190--194} (\bibinfo{year}{2008}).

\bibitem{Fert2017}
\bibinfo{author}{Fert, A.}, \bibinfo{author}{Reyren, N.} \&
  \bibinfo{author}{Cros, V.}
\newblock \bibinfo{title}{{Magnetic skyrmions: advances in physics and
  potential applications}}.
\newblock \emph{\bibinfo{journal}{Nature Reviews Materials}}
  \textbf{\bibinfo{volume}{2}}, \bibinfo{pages}{17031} (\bibinfo{year}{2017}).

\bibitem{Tomasello2014}
\bibinfo{author}{Tomasello, R.} \emph{et~al.}
\newblock \bibinfo{title}{{A strategy for the design of skyrmion racetrack
  memories}}.
\newblock \emph{\bibinfo{journal}{Scientific Reports}}
  \textbf{\bibinfo{volume}{4}}, \bibinfo{pages}{6784} (\bibinfo{year}{2014}).

\bibitem{Luo2018}
\bibinfo{author}{Luo, S.} \emph{et~al.}
\newblock \bibinfo{title}{{Reconfigurable Skyrmion Logic Gates}}.
\newblock \emph{\bibinfo{journal}{Nano Letters}} \textbf{\bibinfo{volume}{18}},
  \bibinfo{pages}{1180--1184} (\bibinfo{year}{2018}).

\bibitem{Li2017}
\bibinfo{author}{Li, S.} \emph{et~al.}
\newblock \bibinfo{title}{{Magnetic skyrmion-based artificial neuron device}}.
\newblock \emph{\bibinfo{journal}{Nanotechnology}}
  \textbf{\bibinfo{volume}{28}}, \bibinfo{pages}{31LT01}
  (\bibinfo{year}{2017}).

\bibitem{Pinna2018}
\bibinfo{author}{Pinna, D.} \emph{et~al.}
\newblock \bibinfo{title}{{Skyrmion Gas Manipulation for Probabilistic
  Computing}}.
\newblock \emph{\bibinfo{journal}{Physical Review Applied}}
  \textbf{\bibinfo{volume}{9}}, \bibinfo{pages}{064018} (\bibinfo{year}{2018}).

\bibitem{Jiang2016}
\bibinfo{author}{Jiang, W.} \emph{et~al.}
\newblock \bibinfo{title}{{Direct observation of the skyrmion Hall effect}}.
\newblock \emph{\bibinfo{journal}{Nature Physics}}
  \textbf{\bibinfo{volume}{13}}, \bibinfo{pages}{162--169}
  (\bibinfo{year}{2016}).

\bibitem{Litzius2016}
\bibinfo{author}{Litzius, K.} \emph{et~al.}
\newblock \bibinfo{title}{{Skyrmion Hall effect revealed by direct
  time-resolved X-ray microscopy}}.
\newblock \emph{\bibinfo{journal}{Nature Physics}}
  \textbf{\bibinfo{volume}{13}}, \bibinfo{pages}{170--175}
  (\bibinfo{year}{2016}).

\bibitem{Reichhardt2016}
\bibinfo{author}{Reichhardt, C.} \& \bibinfo{author}{{Olson Reichhardt}, C.~J.}
\newblock \bibinfo{title}{{Noise fluctuations and drive dependence of the
  skyrmion Hall effect in disordered systems}}.
\newblock \emph{\bibinfo{journal}{New Journal of Physics}}
  \textbf{\bibinfo{volume}{18}}, \bibinfo{pages}{095005}
  (\bibinfo{year}{2016}).

\bibitem{Iwasaki2013b}
\bibinfo{author}{Iwasaki, J.}, \bibinfo{author}{Mochizuki, M.} \&
  \bibinfo{author}{Nagaosa, N.}
\newblock \bibinfo{title}{{Universal current-velocity relation of skyrmion
  motion in chiral magnets}}.
\newblock \emph{\bibinfo{journal}{Nature Communications}}
  \textbf{\bibinfo{volume}{4}}, \bibinfo{pages}{1463} (\bibinfo{year}{2013}).

\bibitem{Hrabec2017}
\bibinfo{author}{Hrabec, A.} \emph{et~al.}
\newblock \bibinfo{title}{{Current-induced skyrmion generation and dynamics in
  symmetric bilayers}}.
\newblock \emph{\bibinfo{journal}{Nature Communications}}
  \textbf{\bibinfo{volume}{8}}, \bibinfo{pages}{15765} (\bibinfo{year}{2017}).

\bibitem{Juge2019}
\bibinfo{author}{Juge, R.} \emph{et~al.}
\newblock \bibinfo{title}{{Current-Driven Skyrmion Dynamics and Drive-Dependent
  Skyrmion Hall Effect in an Ultrathin Film}}.
\newblock \emph{\bibinfo{journal}{Physical Review Applied}}
  \textbf{\bibinfo{volume}{12}}, \bibinfo{pages}{044007}
  (\bibinfo{year}{2019}).

\bibitem{Woo2018}
\bibinfo{author}{Woo, S.} \emph{et~al.}
\newblock \bibinfo{title}{{Current-driven dynamics and inhibition of the
  skyrmion Hall effect of ferrimagnetic skyrmions in GdFeCo films}}.
\newblock \emph{\bibinfo{journal}{Nature Communications}}
  \textbf{\bibinfo{volume}{9}}, \bibinfo{pages}{959} (\bibinfo{year}{2018}).

\bibitem{Dohi2019}
\bibinfo{author}{Dohi, T.}, \bibinfo{author}{DuttaGupta, S.},
  \bibinfo{author}{Fukami, S.} \& \bibinfo{author}{Ohno, H.}
\newblock \bibinfo{title}{{Formation and current-induced motion of synthetic
  antiferromagnetic skyrmion bubbles}}.
\newblock \emph{\bibinfo{journal}{Nature Communications}}
  \textbf{\bibinfo{volume}{10}}, \bibinfo{pages}{5153} (\bibinfo{year}{2019}).

\bibitem{Hirata2019}
\bibinfo{author}{Hirata, Y.} \emph{et~al.}
\newblock \bibinfo{title}{{Vanishing skyrmion Hall effect at the angular
  momentum compensation temperature of a ferrimagnet}}.
\newblock \emph{\bibinfo{journal}{Nature Nanotechnology}}
  \textbf{\bibinfo{volume}{14}}, \bibinfo{pages}{232--236}
  (\bibinfo{year}{2019}).

\bibitem{Zhang2016a}
\bibinfo{author}{Zhang, X.}, \bibinfo{author}{Zhou, Y.} \&
  \bibinfo{author}{Ezawa, M.}
\newblock \bibinfo{title}{Antiferromagnetic skyrmion: Stability, creation and
  manipulation}.
\newblock \emph{\bibinfo{journal}{Scientific Reports}}
  \textbf{\bibinfo{volume}{6}}, \bibinfo{pages}{24795} (\bibinfo{year}{2016}).

\bibitem{Zhang2016b}
\bibinfo{author}{Zhang, X.}, \bibinfo{author}{Zhou, Y.} \&
  \bibinfo{author}{Ezawa, M.}
\newblock \bibinfo{title}{Magnetic bilayer-skyrmions without skyrmion hall
  effect}.
\newblock \emph{\bibinfo{journal}{Nature Communications}}
  \textbf{\bibinfo{volume}{7}}, \bibinfo{pages}{10293} (\bibinfo{year}{2016}).

\bibitem{Zeissler2020}
\bibinfo{author}{Zeissler, K.} \emph{et~al.}
\newblock \bibinfo{title}{{Diameter-independent skyrmion Hall angle observed in
  chiral magnetic multilayers}}.
\newblock \emph{\bibinfo{journal}{Nature Communications}}
  \textbf{\bibinfo{volume}{11}}, \bibinfo{pages}{428} (\bibinfo{year}{2020}).

\bibitem{Kim2017}
\bibinfo{author}{Kim, J.-V.} \& \bibinfo{author}{Yoo, M.-W.}
\newblock \bibinfo{title}{{Current-driven skyrmion dynamics in disordered
  films}}.
\newblock \emph{\bibinfo{journal}{Applied Physics Letters}}
  \textbf{\bibinfo{volume}{110}}, \bibinfo{pages}{132404}
  (\bibinfo{year}{2017}).

\bibitem{Legrand2017}
\bibinfo{author}{Legrand, W.} \emph{et~al.}
\newblock \bibinfo{title}{{Room-Temperature Current-Induced Generation and
  Motion of sub-100 nm Skyrmions}}.
\newblock \emph{\bibinfo{journal}{Nano Letters}} \textbf{\bibinfo{volume}{17}},
  \bibinfo{pages}{2703--2712} (\bibinfo{year}{2017}).

\bibitem{Iwasaki2013}
\bibinfo{author}{Iwasaki, J.}, \bibinfo{author}{Mochizuki, M.} \&
  \bibinfo{author}{Nagaosa, N.}
\newblock \bibinfo{title}{{Current-induced skyrmion dynamics in constricted
  geometries}}.
\newblock \emph{\bibinfo{journal}{Nature Nanotechnology}}
  \textbf{\bibinfo{volume}{8}}, \bibinfo{pages}{742--747}
  (\bibinfo{year}{2013}).

\bibitem{Yoo2017}
\bibinfo{author}{Yoo, M.-W.}, \bibinfo{author}{Cros, V.} \&
  \bibinfo{author}{Kim, J.-V.}
\newblock \bibinfo{title}{{Current-driven skyrmion expulsion from magnetic
  nanostrips}}.
\newblock \emph{\bibinfo{journal}{Physical Review B}}
  \textbf{\bibinfo{volume}{95}}, \bibinfo{pages}{184423}
  (\bibinfo{year}{2017}).

\bibitem{Woo2017}
\bibinfo{author}{Woo, S.} \emph{et~al.}
\newblock \bibinfo{title}{{Spin-orbit torque-driven skyrmion dynamics revealed
  by time-resolved X-ray microscopy}}.
\newblock \emph{\bibinfo{journal}{Nature Communications}}
  \textbf{\bibinfo{volume}{8}}, \bibinfo{pages}{15573} (\bibinfo{year}{2017}).

\bibitem{Emori2015}
\bibinfo{author}{Emori, S.}, \bibinfo{author}{Umachi, C.~K.},
  \bibinfo{author}{Bono, D.~C.} \& \bibinfo{author}{Beach, G.~S.}
\newblock \bibinfo{title}{{Generalized analysis of thermally activated
  domain-wall motion in Co/Pt multilayers}}.
\newblock \emph{\bibinfo{journal}{Journal of Magnetism and Magnetic Materials}}
  \textbf{\bibinfo{volume}{378}}, \bibinfo{pages}{98--106}
  (\bibinfo{year}{2015}).

\bibitem{Du2015a}
\bibinfo{author}{Du, H.} \emph{et~al.}
\newblock \bibinfo{title}{{Edge-mediated skyrmion chain and its collective
  dynamics in a confined geometry}}.
\newblock \emph{\bibinfo{journal}{Nature Communications}}
  \textbf{\bibinfo{volume}{6}}, \bibinfo{pages}{8504} (\bibinfo{year}{2015}).

\bibitem{Lai2017}
\bibinfo{author}{Lai, P.} \emph{et~al.}
\newblock \bibinfo{title}{{An Improved Racetrack Structure for Transporting a
  Skyrmion}}.
\newblock \emph{\bibinfo{journal}{Scientific Reports}}
  \textbf{\bibinfo{volume}{7}}, \bibinfo{pages}{45330} (\bibinfo{year}{2017}).

\bibitem{Bessarab2018}
\bibinfo{author}{Bessarab, P.~F.} \emph{et~al.}
\newblock \bibinfo{title}{{Lifetime of racetrack skyrmions}}.
\newblock \emph{\bibinfo{journal}{Scientific Reports}}
  \textbf{\bibinfo{volume}{8}}, \bibinfo{pages}{3433} (\bibinfo{year}{2018}).

\bibitem{Thiele1973}
\bibinfo{author}{Thiele, A.~A.}
\newblock \bibinfo{title}{{Steady-State Motion of Magnetic Domains}}.
\newblock \emph{\bibinfo{journal}{Physical Review Letters}}
  \textbf{\bibinfo{volume}{30}}, \bibinfo{pages}{230--233}
  (\bibinfo{year}{1973}).

\bibitem{Locatelli2013}
\bibinfo{author}{Locatelli, N.}, \bibinfo{author}{Cros, V.} \&
  \bibinfo{author}{Grollier, J.}
\newblock \bibinfo{title}{{Spin-torque building blocks}}.
\newblock \emph{\bibinfo{journal}{Nature Materials}}
  \textbf{\bibinfo{volume}{13}}, \bibinfo{pages}{11--20}
  (\bibinfo{year}{2013}).

\bibitem{Knoester2014}
\bibinfo{author}{Knoester, M.~E.}, \bibinfo{author}{Sinova, J.} \&
  \bibinfo{author}{Duine, R.~A.}
\newblock \bibinfo{title}{{Phenomenology of current-skyrmion interactions in
  thin films with perpendicular magnetic anisotropy}}.
\newblock \emph{\bibinfo{journal}{Physical Review B}}
  \textbf{\bibinfo{volume}{89}}, \bibinfo{pages}{064425}
  (\bibinfo{year}{2014}).

\bibitem{Vansteenkiste2014}
\bibinfo{author}{Vansteenkiste, A.} \emph{et~al.}
\newblock \bibinfo{title}{{The design and verification of MuMax3}}.
\newblock \emph{\bibinfo{journal}{AIP Advances}} \textbf{\bibinfo{volume}{4}},
  \bibinfo{pages}{107133} (\bibinfo{year}{2014}).

\bibitem{Reichhardt2015}
\bibinfo{author}{Reichhardt, C.}, \bibinfo{author}{Ray, D.} \&
  \bibinfo{author}{Reichhardt, C. J.~O.}
\newblock \bibinfo{title}{{Collective Transport Properties of Driven Skyrmions
  with Random Disorder}}.
\newblock \emph{\bibinfo{journal}{Physical Review Letters}}
  \textbf{\bibinfo{volume}{114}}, \bibinfo{pages}{217202}
  (\bibinfo{year}{2015}).

\bibitem{Parkin2015}
\bibinfo{author}{Parkin, S.} \& \bibinfo{author}{Yang, S.-H.}
\newblock \bibinfo{title}{{Memory on the racetrack}}.
\newblock \emph{\bibinfo{journal}{Nature Nanotechnology}}
  \textbf{\bibinfo{volume}{10}}, \bibinfo{pages}{195--198}
  (\bibinfo{year}{2015}).

\bibitem{Zhao2012}
\bibinfo{author}{Zhao, W.~S.} \emph{et~al.}
\newblock \bibinfo{title}{{Magnetic domain-wall racetrack memory for high
  density and fast data storage}}.
\newblock In \emph{\bibinfo{booktitle}{11th International Conference on
  Solid-State and Integrated Circuit Technology}}, \bibinfo{pages}{1--4}
  (\bibinfo{year}{2012}).

\bibitem{Ho2019}
\bibinfo{author}{Ho, P.} \emph{et~al.}
\newblock \bibinfo{title}{{Geometrically Tailored Skyrmions at Zero Magnetic
  Field in Multilayered Nanostructures}}.
\newblock \emph{\bibinfo{journal}{Physical Review Applied}}
  \textbf{\bibinfo{volume}{11}}, \bibinfo{pages}{024064}
  (\bibinfo{year}{2019}).

\bibitem{He2016}
\bibinfo{author}{He, S.} \emph{et~al.}
\newblock \bibinfo{title}{{A Versatile Rotary-Stage High Frequency Probe
  Station for Studying Magnetic Films and Devices}}.
\newblock \emph{\bibinfo{journal}{Review of Scientific Instruments}}
  \textbf{\bibinfo{volume}{87}}, \bibinfo{pages}{074704}
  (\bibinfo{year}{2016}).

\bibitem{Lourembam2018}
\bibinfo{author}{Lourembam, J.}, \bibinfo{author}{Ghosh, A.} \&
  \bibinfo{author}{Zeng, M.}
\newblock \bibinfo{title}{{Thickness-Dependent Perpendicular Magnetic
  Anisotropy and Gilbert Damping in Hf/Co20Fe60B20/MgO Heterostructures}}.
\newblock \emph{\bibinfo{journal}{Physical Review Applied}}
  \textbf{\bibinfo{volume}{10}}, \bibinfo{pages}{044057}
  (\bibinfo{year}{2018}).

\bibitem{Kittel1948}
\bibinfo{author}{Kittel, C.}
\newblock \bibinfo{title}{{On the Theory of Ferromagnetic Resonance
  Absorption}}.
\newblock \emph{\bibinfo{journal}{Physical Review}}
  \textbf{\bibinfo{volume}{73}}, \bibinfo{pages}{155} (\bibinfo{year}{1948}).

\bibitem{Celinski1991}
\bibinfo{author}{Celinski, Z.} \& \bibinfo{author}{Heinrich, B.}
\newblock \bibinfo{title}{{Ferromagnetic resonance linewidth of Fe ultrathin
  films grown on a bcc Cu substrate}}.
\newblock \emph{\bibinfo{journal}{Journal of Applied Physics}}
  \textbf{\bibinfo{volume}{70}}, \bibinfo{pages}{5935} (\bibinfo{year}{1991}).

\bibitem{Tan2020}
\bibinfo{author}{Tan, A. K.~C.} \emph{et~al.}
\newblock \bibinfo{title}{Skyrmion generation from irreversible fission of
  stripes in chiral multilayer films}.
\newblock \emph{\bibinfo{journal}{Phys. Rev. Materials}}
  \textbf{\bibinfo{volume}{4}}, \bibinfo{pages}{114419} (\bibinfo{year}{2020}).

\end{thebibliography}

\newpage
\widetext

\makeatletter
\makeatletter\@addtoreset{paragraph}{section}\makeatother
\makeatletter\def\p@paragraph{}\makeatother
\makeatletter\def\l@section{\@dottedtocline{1}{0.6em}{1.5em}}\makeatother
\makeatletter\def\l@paragraph{\@dottedtocline{4}{1.5em}{1.8em}}\makeatother
\makeatletter\def\l@figure{\@dottedtocline{1}{0.6em}{1.8em}}\makeatother

\renewcommand{\thefigure}{S\arabic{figure}}
\renewcommand\theHfigure{S\arabic{figure}}
\renewcommand{\theequation}{S\arabic{equation}}
\renewcommand{\thetable}{S\arabic{table}}
\renewcommand{\thesection}{S\arabic{section}}
\setcounter{secnumdepth}{4}
\setcounter{equation}{0}
\setcounter{section}{0}
\setcounter{figure}{0}
\setcounter{table}{0}

\titleformat{\section}{\large\bfseries\scshape\filcenter}{\thesection.}{1em}{#1}[{\titlerule[0.5pt]}]
\titlespacing*{\section}{0pt}{1ex}{1ex}
\titleformat{\subsection}{\bfseries\sffamily}{\thessubsection.}{0em}{#1}
\titlespacing{\subsection}{0pt}{0.5ex}{0.5ex}
\titleformat{\paragraph}[runin]{\sffamily\bfseries}{}{-1.2em}{#1.}
\titlespacing*{\paragraph}{1.25em}{2ex}{0.4em}[] 

\begin{center}
	\Large\textbf{Supplementary Information for \\}
	\large\textbf{Visualizing the Strongly Reshaped Skyrmion Hall Effect \\ in Multilayer Wire Devices\smallskip{}}
\end{center}

\linespread{2}
\setlength{\parskip}{3ex plus0.2ex minus0.2ex}

\linespread{1.25}
\setlength{\parskip}{1ex plus0.2ex minus0.2ex}

\section{Stack Structure \& Magnetic Properties\label{sec:MagProps}}

\begin{figure}[h]
\begin{centering}
\includegraphics[width=4.5in]{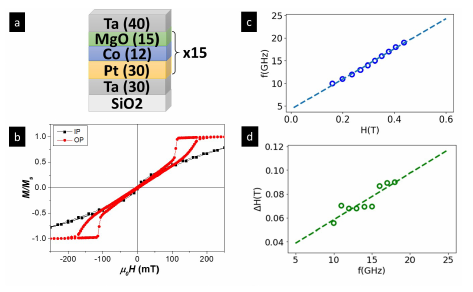}
\end{centering}
\noindent \caption[Hysteresis Loop \& FMR ]{\textbf{Stack Structure and Magnetic Parameters.}
\textbf{(a)} Schematic representation of the multilayer stack containing 15 repeats of Pt(3)/Co(1.2)/MgO(1.5) (thickness in nm in parentheses). \textbf{(b)} Out-of-plane (OP, red) and in-plane (IP, black) magnetization hysteresis loops. \textbf{(c-d)} Fits to the ferromagnetic resonance (FMR) peak positions (c) and linewidths (d) for the determination of gyromagnetic ratio $\gamma$ and Gilbert damping parameter $\alpha$ respectively.
\label{fig:ExptMHLoop}}
\end{figure}

\paragraph{Stack Structure}
A multilayer film with 15 repeats of Pt(3)/Co(0.9-1.4)/MgO(1.5) (norminal thicknesses in nm in parentheses) was deposited on 8'' thermally oxidized Si wafer using Singulus Timaris\texttrademark . Coupon samples were precisely diced at positions of interest i.e. Co thickness of 1.2~nm, for subsequent characterization and device fabrication. The expected Co thickness variation along the wedge for a coupon size of $1\times 1$~cm is negligibly small ($< 0.02$~nm). The complete stack structure, illustrated schematically in \ref{fig:ExptMHLoop}a consists of a Ta (3~nm) underlayer for adhesion and Pt texture and Ta (4~nm) cap for protection against oxidation.

\paragraph{Magnetic Parameters}
The [Pt(3)/Co(1.2)/MgO(1.5)]$_{15}$ multilayer thin film has an effective OP anisotropy ($K_{\rm eff}$) of 0.17~MJ/m$^3$ and saturation magnetization ($M_{\rm s}$) of 1.1~MA/m, as derived from out-of-plane (OP) and in-plane (IP) magnetic hysteresis measurements using vibrating sample magnetometry (\ref{fig:ExptMHLoop}b, details in Methods). The interfacial Dzyaloshinskii-Moriya interaction (DMI, $D$) of 1.6~mJ/m$^{2}$ and exchange stiffness ($A$) of 24~pJ/m were determined using an established combination of experiments and simulations. $\chi$$^{2}$-fits were performed to the zero field domain periodicity (340~nm) measured by magnetic force microscopy (MFM) against micromagnetic simulation results for a wide range of $D$ and $A$\citep{Soumyanarayanan2017,Ho2019}.

\paragraph{Gilbert Damping}
The Gilbert damping parameter ($\alpha$) was determined to be 0.05 from magnetization dynamics measurements using a home-built broadband vector network analyzer ferromagnetic resonance spectroscopy (VNA-FMR) setup\citep{He2016,Lourembam2018}. The FMR spectra were obtained for external OP magnetic fields of up to 0.55~T over a frequency ($f$) range of 2 to 26~GHz, and fitted using Lorentz absorption and dispersion line-shape functions\citep{Lourembam2018}. To obtain the gyromagnetic ratio $\gamma$ (\ref{fig:ExptMHLoop}c), the FMR resonance peaks
($H$) were fitted using the Kittel equation for OP configuration, defined as\citep{Kittel1948}
\begin{equation}
f=\frac{\mu_{\rm 0}\mid\gamma\mid}{2\pi}(H - M_{\rm eff})
\label{eq:Kittel eq}\end{equation}
where $M_{\rm eff}$ is the effective magnetization. The $\alpha$ (\ref{fig:ExptMHLoop}d) was then determined from the linear fit of the full width half maxima of the peaks ($\Delta H$) vs $f$, defined as\citep{He2016,Celinski1991}
\begin{equation}
\Delta H=\frac{4\pi\alpha}{\mu_{\rm 0}\mid\gamma\mid}f+\Delta H_{\rm 0}
\label{eq:H-fLinear fit}\end{equation}
where $\Delta H_{{\rm 0}}$ represents the inhomogeneous linewidth broadening.

\clearpage{}

\section{Wire Device Fabrication\label{sec:Device Fab}}

\begin{figure}[h]
\begin{centering}
\includegraphics[width=5in]{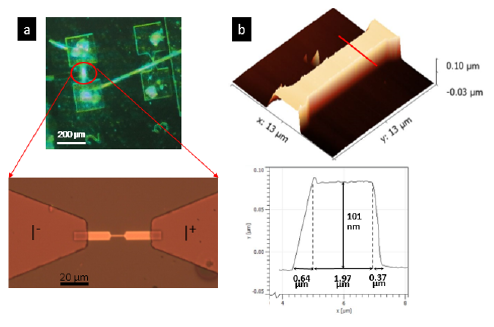}
\end{centering}
\caption[Device fab]{\textbf{Wire Device Fabrication.} 
\textbf{(a)} Optical microscope image of a [Pt(3)/Co(1.2)/MgO(1.5)]$_{15}$ wire device with Ta(5)/Au(100)/Ru(20) (nominal layer thicknesses in nm in parentheses) electrode wire-bonded to the chip carrier. 
\textbf{(b)} Atomic force microscopy 3D image of a representative wire (top) and its cross-sectional height profile (bottom) at the wire position indicated in red.
\label{fig:Devicefab}}\end{figure}

\paragraph{Wire Profile}
Coupon samples consisting of Pt(3)/Co(1.2)/MgO(1.5) multilayer films were patterned with electron beam lithography and etched, as detailed in Methods, to produce wire devices of dimensions $2\,\times\,10\,\mu$m (\ref{fig:Devicefab}a). The Ta/Au/Ru electrodes were subsequently patterned on the wire (see Methods) for wire-bonding to the chip carrier. The etching process was closely monitored with an end-point detector and over-etched by $\sim 10$~nm into the SiO$_2$. Atomic force microscopy of a representative wire (\ref{fig:Devicefab}b) shows a tapered wire profile with intermittent defects along the edges due to incomplete lift-off of resist or side wall re-deposition. The topographic spikes are rather small in height ($\lesssim 5$~nm, i.e. $\lesssim 5\%$ of total stack thickness) and cover $<8\%$ of the wire width. Notably, any surface defects are limited in their detriment to the magnetic contrast and skyrmion dynamics at the edge. Notably, we observe near-unitary skyrmion distribution across wire width with no preferential existence of skyrmions at the edge (see manuscript Fig. 3c).

\clearpage{}

\section{Skyrmion Nucleation \& Properties\label{sec:SkProperties}}

\begin{figure}[h]
\begin{centering}
\includegraphics[width=2.6in]{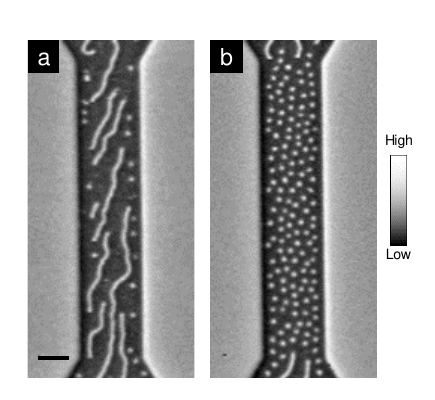}
\end{centering}
\caption[Sk Creation Protocol]{\textbf{Skyrmion Creation Protocol.} 
MFM images (scale bar: 100~nm) of the wire at a representative OP field, after negative saturation, showing \textbf{(a)} initialized magnetic textures consisting of stripes and skyrmions, and \textbf{(b)} an array of skyrmions after following a pulse injection protocol to break up the stripes.\label{fig:SkCreation}}
\end{figure}

\paragraph{Skyrmion Creation}
At lower magnetic fields, stripe domains may be stabilized along with skyrmions. To avoid ambiguity, we established a protocol involving current-induced fission of stripes\citep{Tan2020} to ensure that the wire consists of solely skyrmions for dynamics studies. The protocol involved injecting current pulses with gradually increasing magnitude $J$, and observing MFM images for changes to the magnetic configuration. This enabled the selection of a minimum $J$ for skyrmion creation while mitigating physical device damage or skyrmion annihilation due to heating. The pulsing continued at this $J$ until no skyrmions were created with further pulsing. This protocol works only above a certain magnetic field threshold wherein skyrmions are metastable\citep{Tan2020} -- below which no amount of pulsing could create a skyrmion configuration. Therefore field values below the threshold were not included in this study.

\begin{figure}[h]
\begin{centering}
\includegraphics[width=2.8in]{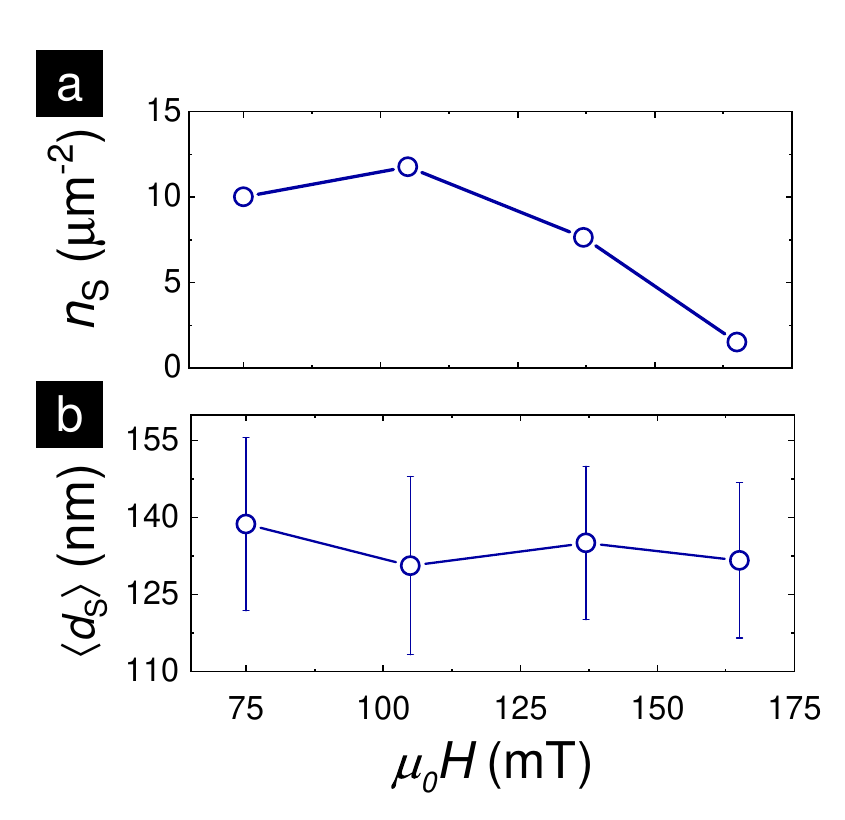}
\end{centering}
\caption[Sk Parameters]{\textbf{Skyrmions Properties.} 
\textbf{(a)} Density $n_{\rm S}$, and \textbf{(b)} field-of-view average size $\langle d_{\rm S}\rangle$ -- of skyrmion configurations as a function of OP field after \emph{ex situ} saturation at -400~mT followed by the skyrmion creation protocol (see above).
\label{fig:Skparameters}}\end{figure}

\paragraph{Skyrmion Properties}
The skyrmion density ($n_{\rm S}$, \ref{fig:Skparameters}a) and field-of-view average size ($\langle d_{\rm S}\rangle$, \ref{fig:Skparameters}b) were measured from MFM images of skyrmion lattice configurations stabilized at various OP applied fields of 75 to 165~mT (detailed in manuscript Fig. 1). The $n_{\rm S}$ peaks at $\sim 13\,\mu$m$^{-2}$ under an OP applied field of 105~mT, followed by a gradual decline to $\sim 2\,\mu$m$^{-2}$ with increasing field. Meanwhile, the $\langle d_{\rm S}\rangle$ shows a monotonic $\sim 6\%$ decrease from 140 to 132~nm over the range of applied fields. The large variance of skyrmion sizes at each field ($\sim 80-200$~nm, see manuscript Fig. 4) enables statistically significant analysis of the size dependence of skyrmion dynamics.
\clearpage{}
\section{Skyrmion Motion Tracking Protocol\label{sec:SkTracking}}

\begin{figure}[h]
\begin{centering}
\includegraphics[width=5in]{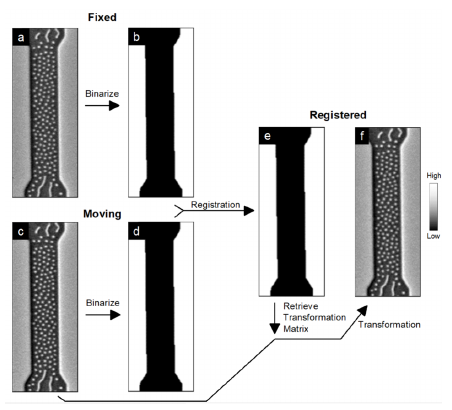}
\end{centering}
\caption[ImageReg]{\textbf{Registration of MFM Images.}
\textbf{(a-b)}Reference MFM image (a) and the corresponding reference binary image (b) of the wire device.
\textbf{(c-d)} Target MFM image (c) and the corresponding target binary image (d). \textbf{(e)} The resulting registered binary image derived from the registration of target (d) to reference (b).
\textbf{(f)} The transformed MFM image of (c) based on the transformation matrix retrieved from (e).
\label{fig:ImageReg}}\end{figure}

\paragraph{Image Correction \& Registration}
The MFM images were initially processed using common scanning probe microscopic (SPM) image correction techniques including plane surface subtraction and row alignment. For the tracking of skyrmion motion, the wire device needs to be imaged at an identical position before and after each applied pulse. Since drifts inherent to SPM cannot be fully eradicated in experiments, image registration -- i.e. post alignment of the image to a reference -- is required to correct for slight position shifts. One means of registration is to align the device topography in each dataset. However, topographic artifacts due to fluctuations in feedback and scan speed deem such a technique unreliable in practice. Instead we adopt a less rigorous approach by using the MFM channel -- which is less susceptible to scan speed issues and feedback fluctuations. To remove any potential registration bias from the domain patterns, the MFM images from the reference (fixed) and the target (moving) dataset are binarized and any hole present is filled up (\ref{fig:ImageReg}a-d). Following this, the transformation matrix ($TF$) aligning the moving to the fixed binary data (\ref{fig:ImageReg}e) is retrieved and applied to the moving MFM data (\ref{fig:ImageReg}c) to obtain a registered MFM data (\ref{fig:ImageReg}f).

\begin{figure}[h]
\begin{centering}
\includegraphics[width=6.5in]{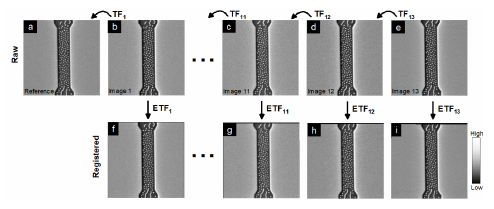}
\end{centering}
\caption[Cumulative Reg]{\textbf{Cumulative Registration of Consecutively Acquired MFM Images.}
\textbf{(a-e)} Set of consecutively acquired MFM images where $TF_{x}$ is the transformation matrix aligning image $x$ to image $(x-1)$.
\textbf{(f-i)} Registered images of (b-e) aligned to reference image (a) where $ETF_{x}$ is the effective transformation matrix defined in \ref{eq:ETF}.
\label{fig:CumulativeReg}}\end{figure}

\paragraph{Cumulative Registration}
The intensity-based image registration was implemented using the Image Processing Toolbox in MATLAB$^{\circledR}$ assuming a 2D affine transformation. Since the drifts from consecutive scans are relatively minor - evident in \ref{fig:ImageReg}, a negligibly small downward translation between the moving and fixed image, registrations are carried out on consecutive images instead of the reference dataset. The effective $TF$ ($ETF$) aligning each dataset to the reference is then obtained by multiplying $TF$s aligning prior images cumulatively, defined as
\begin{equation}
ETF_{x}=\overset{x}{\prod_{1}}TF_{i}
\label{eq:ETF}\end{equation}
As shown in \ref{fig:CumulativeReg}, Image 1 aligns to the Reference via $TF_1$; Image 2 aligns to the Reference via $TF_2$ ($TF$ between 1st and 2nd image) $\times TF_1$; Image 3 aligns to the Reference via $TF_3\,\times\,TF_2\,\times TF_1$ and so on. This registration protocol rigorously eliminates any erroneous registration results especially when a long time has elapsed between the acquisition of the reference and target MFM image.

\begin{figure}[h]
\begin{centering}
\includegraphics[width=3in]{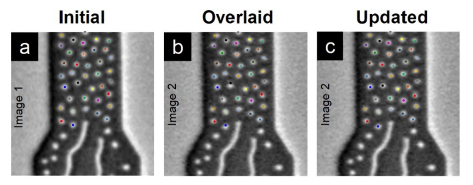}
\end{centering}
\caption[Sk Tagging and Tracking]{\textbf{Skyrmion Tracking.}
Initial skyrmion positions (shown as dots) overlaid on MFM images acquired \textbf{(a)} before pulse and \textbf{(b)} after pulse.
\textbf{(c)} Final skyrmion positions overlaid on MFM image acquired after pulse. Each skyrmion is uniquely marked with a colored marker.
\label{fig:SkTrack}}\end{figure}

\paragraph{Skyrmion Tagging and Tracking}
Upon generating skyrmions in the wire device (skyrmion nucleation protocol described in \ref{sec:SkProperties}), the skyrmions in the MFM image are identified and uniquely tagged. The identification of skyrmions is noticeably easier here as it does not involve differentiating stripes and skyrmions. To aid tracking, the previous skyrmion tags are overlaid on the next MFM image obtained after a pulse is injected. The skyrmion tags are updated from their previous positions by systematically assigning each tag to their respective new skyrmion position. It is crucial that the assignment of new positions follows a systematic top-down (or bottom-up) direction, ensuring each tag is assigned to the nearest skyrmion as much as possible. The tracking process is repeated for every pulse. The tracking protocol is rigorous only if the number of skyrmions remain nearly unchanged during pulsing. Importantly, any large fluctuation in numbers (skyrmion annihilation/creation) indicates a high likelihood of device damage due to substantial current-induced heating. Consequently, experiments were terminated or discarded from analysis in cases where skyrmion numbers showed drastic changes.

\clearpage{}

\section{Quantifying Skyrmion Dynamics Parameters\label{sec:SkSize}}

\paragraph{Retrieval of Skyrmion Size}
To reduce bias arising from the choice of fit window size, we adopt a two-step fit process to obtain the $d_{\rm S}$. The first fit iteration allows the estimation of the fitting parameters which are used to optimize the fitting window length to $1.5\times$ of the estimated $d_{\rm S}$. The second iteration uses the optimized window for fitting, with the estimated values as the initial guess. In both steps, the method of least squares was used to fit a Gaussian function, $f(x_{i},y_{j},\beta)$ to the observed signal, $z(x_{i},y_{j})$. To place the fit emphasis on the vicinity of the skyrmion, with radially decaying importance towards the corners of the fitting window, the residual, $r_{ij}$ is therefore weighted by another Gaussian function, $f_{\rm weight}(x_{i},y_{j},\beta)$, defined as
\begin{equation}
r_{ij} = [f(x_i,y_j,\beta)-z(x_i,y_j)]\cdot f_{\rm weight}(x_i,y_j,\beta)
\label{eq:residual}\end{equation}
\begin{equation}
f(x_i,y_j,\beta) = a\cdot\exp\left\{-\frac{(x_i- x_0)^2 + (y_j-y_0)^2}{2\sigma^2}\right\} + b
\label{eq:Gaussian}\end{equation}
\begin{equation}
f_{\rm weight}(x_i,y_j,\beta) = \exp\left\{- \frac{(x_i-x_0)^2 + (y_j-y_0)^2}{2\sigma^2}\right\}
\label{eq:Gaussian weight}\end{equation}
for set of fit parameters, $\beta=(a,b,x_0,y_0,\sigma)$. The $d_{\rm S}$ is hence defined by the linewidth of the fitted Gaussian, $f(\beta)$, given as $2\sqrt{(2\ln2)}\sigma$.

\begin{figure}[h]
\begin{centering}
\includegraphics[width=3in]{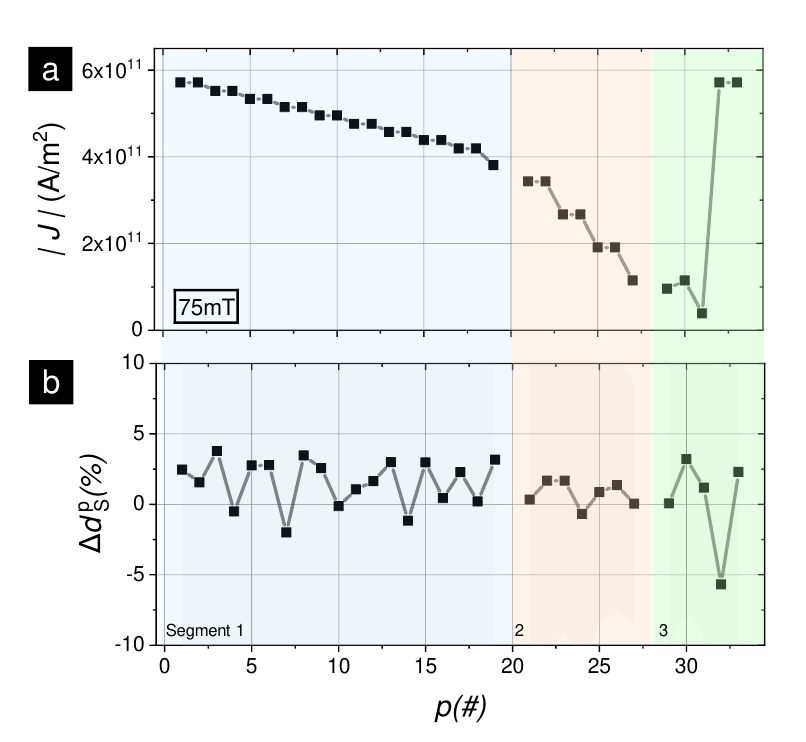}
\end{centering}
\caption[Size Change with pulse]{\textbf{Change in Skyrmion Size.}
\textbf{(a)} Magnitude of current density, $|J|$ and \textbf{(b)} corresponding change in skyrmion size, $\Delta d_{\rm S}^{\,p}$ for each applied pulse, $p$ at OP field of 75mT. \label{fig:Size-Pulse}}\end{figure}

\paragraph{Current-Induced Size Changes}
Due to the raster scanning nature of the MFM technique, images of the magnetic configuration can be acquired only before or after a current pulse is applied. The accuracy of the size analysis will therefore depend on the perturbation of $d_{\rm S}$ in response to the applied current pulse. We quantify this by recording the skyrmion size change, $\Delta d_{{\rm S}_i}^{\,p}$ after each pulse $p$, and quantifying the average change, $\Delta d_{\rm S}^{\,p}$ given as:
\begin{equation}
\Delta d_{{\rm S}_i}^{\,p}(\%)=\frac{d_{{\rm S}_i}^{\,p} - d_{{\rm S}_i}^{\,p-1}}
{d_{{\rm S}_i}^{\,p-1}},\;
{\rm S}_i\text{ is the }i^{\rm th}\text{ skyrmion, and }
\label{eq:indiv size change}\end{equation}
\begin{equation}
\Delta d_{\rm S}^{\,p}(\%) =
\frac{\sum_i^N \Delta d_{{\rm S}_i}^{\,p}}{N},\;
N\text{ is the total number of skyrmions.}
\label{eq:ave size change}\end{equation}
We find that after each pulse $p$, $\mid\Delta d_{\rm S}^{\,p}\mid$ is $\lesssim 5\%$ for $|J|$ up to $6.0\times10^{11}$~A/m$^2$ (\ref{fig:Size-Pulse}). This indicates that the size analysis in manuscript Fig. 4 is valid to $\pm 10$~nm for larger $d_{\rm S}$ ($<200$~nm) and to $\pm 5$~nm for smaller $d_{\rm S}$ ($<100$~nm).

\paragraph{Dynamics Analysis}
The parameters describing skyrmion motion -- skyrmion velocity, $v_{\rm S}$ and deflection angle $\theta_{\rm S}$ -- are characterized with reference to the direction of $J$ (see manuscript Fig. 2a). The $v_{\rm S}$ is given by the ratio of displacement of the skyrmion after the pulse injection to the effective pulse duration of 20~ns. The $\langle v_{\rm S}\rangle$ detailed in manuscript Fig. 2 only includes skyrmions that are in motion, wherein any positional change of $<1$ pixel ($\sim 50$~nm) is considered as static. Meanwhile, the $\theta_{\rm S}$ is wrapped in the range of $-90^\circ$ to $90^\circ$, with $0^\circ$ being the direction of $J$. Specifically, in instances of $\theta_{\rm S}$ less than $-90^\circ$ or more than $90^\circ$, a value of $180^\circ$ is added or subtracted, respectively. This analysis was employed throughout the study with the exception of the plastic flow regime analyses (detailed in manuscript Fig. 3 and 4) where this angular projection is not valid -- only skyrmions that move with $J$ are considered. The data for $v_{\rm S}$ and $\theta_{\rm S}$ analyses are binned and only bins with more than 5~skyrmions ($N_{\rm cutoff} > 5$) are considered for averaging. The $N_{\rm cutoff}$ is reduced to 1 for the skyrmion size analysis in manuscript Fig. 4 as the dataset is further analyzed for $J$ dependence. The robustness of the analysis is addressed in \ref{sec:BinningChecks}.

\clearpage{}

\section{Additional Data for Skyrmion Flow Dynamics\label{sec:105mT}}

\begin{figure}[h]
\begin{centering}
\includegraphics[width=2.75in]{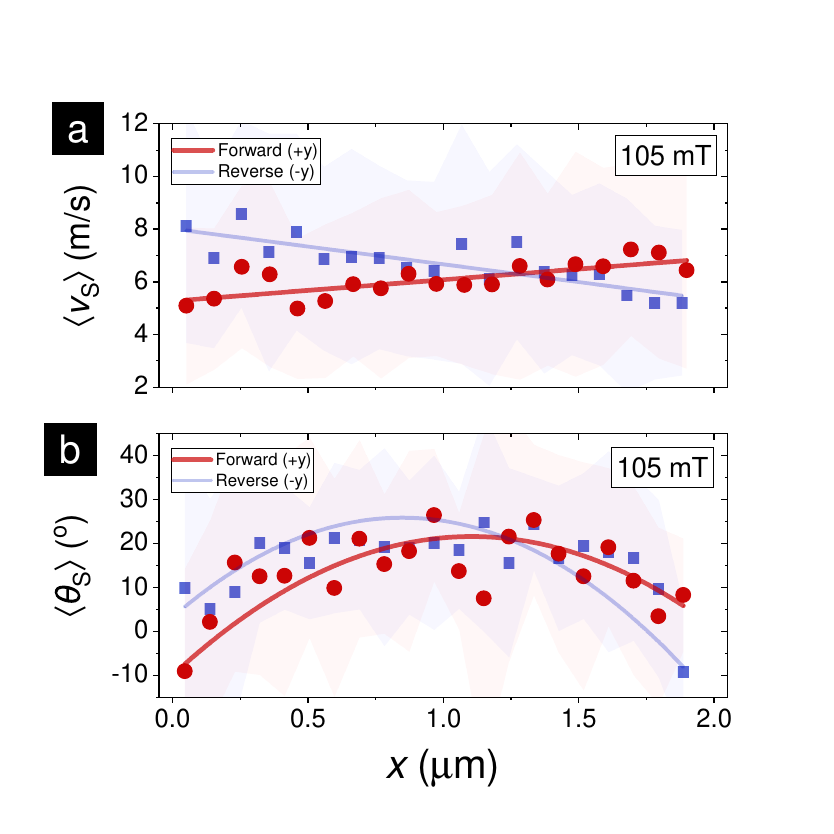}
\end{centering}
\caption[Edge Effect at 105mT]{\textbf{Confinement Effects on Skyrmion Flow Dynamics}. At $\mu_0 H \simeq 105$~mT, \textbf{(a)} the average velocity $\langle v_{\rm S}(x)\rangle$ and \textbf{(b)} angular deflection $\langle\theta_{\rm S}(x)\rangle$ for skyrmions in each $x$-bin for forward ($J \parallel +\hat{y}$, blue) and reverse ($J \parallel -\hat{y}$, red) motion. Solid lines are guides-to-the-eye, while shaded regions represent the standard deviation. \label{fig:Edge Effect 105mT}}\end{figure}

\paragraph{Edge Dependence of Velocity and SkHE}
Analogous to manuscript Fig. 3d-e (75~mT), \ref{fig:Edge Effect 105mT} shows the variation of $\langle v_{\rm S}(x)\rangle$ and $\langle\theta_{\rm S}(x)\rangle$ with transverse position $x$ at 105~mT along the forward ($J \parallel +\hat{y}$) and reverse ($J \parallel -\hat{y}$) directions. Consistent with 75~mT observations, there is a linear decrease in $\langle v_{\rm S}(x)\rangle$ by $\sim30\%$ at the left edge of the wire in the forward motion (\ref{fig:Edge Effect 105mT}). Additionally, for both OP fields of 75~mT (manuscript Fig. 3e) and 105~mT (\ref{fig:Edge Effect 105mT}b), the $\langle\theta_{\rm S}(x)\rangle$ increases from negative at $x = 0\,\mu$m to its maximally positive value at $x \sim 1\,\mu$m, and then decreases as $x$ approaches $2\,\mu$m. The same variation for $\langle v_{\rm S}(x)\rangle$ and $\langle\theta_{\rm S}(x)\rangle$ are observed for both directions.

\begin{figure}[h]
\begin{centering}
\includegraphics[width=2.5in]{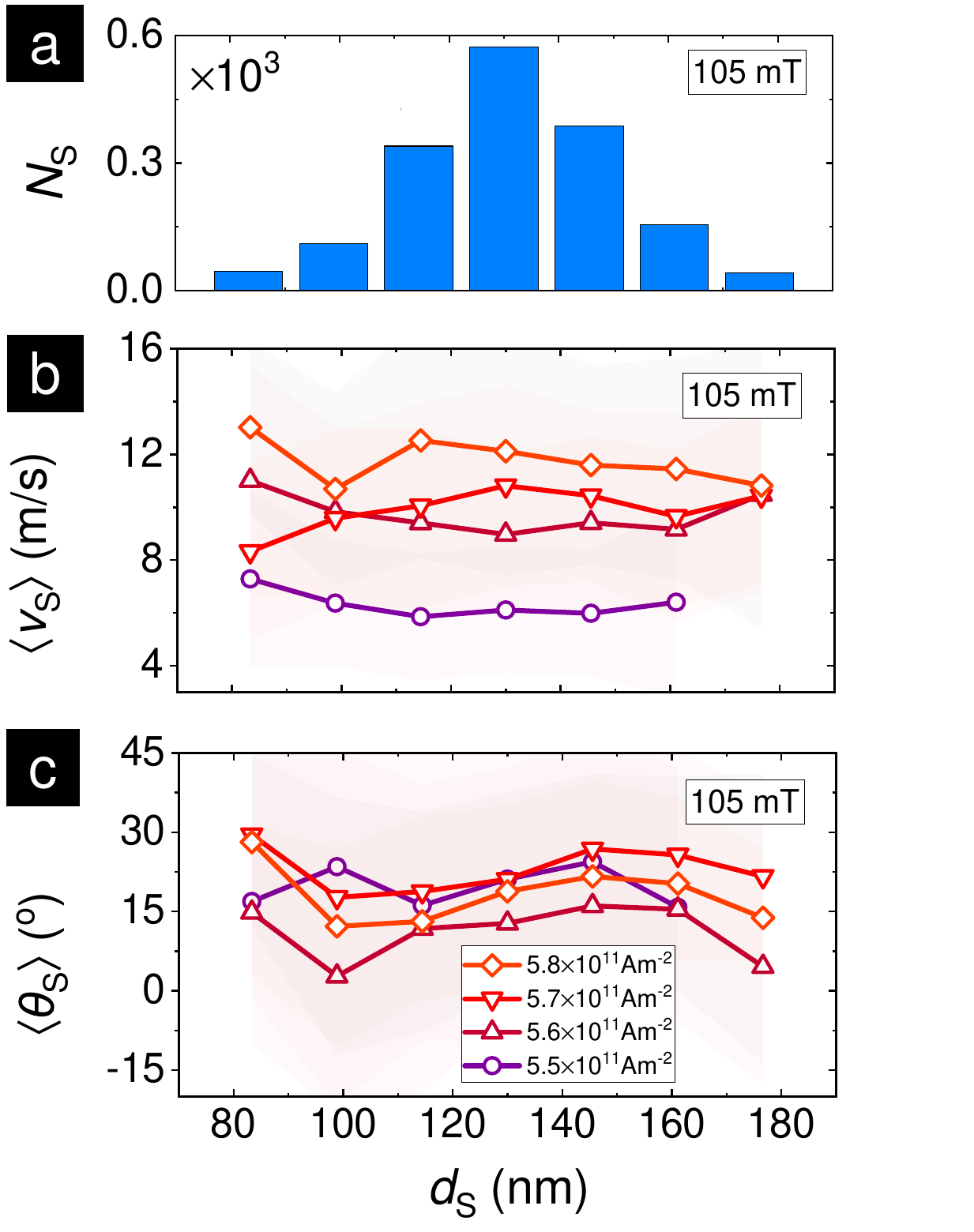}
\end{centering}
\caption[Skyrmion Size Effect at 105mT]{\textbf{Skyrmion Size Effect on Skyrmion Flow Dynamics.} At $\mu_0 H \simeq 105$~mT, \textbf{(a)} binned histogram distribution of skyrmions in the plastic flow regime, $N_{\rm S}$, based on their size, $d_{\rm S}$ -- which varies over 80-180~nm. \textbf{(b-c)} Average velocity $\langle v_{\rm S}\rangle$ (b) and angular deflection $\langle\theta_{\rm S}\rangle$ (c) for skyrmions in each $d_{\rm S}$-bin across the plastic flow regime ($J= (5.5-5.8)\times 10^{11}$~A/m$^2$). Shaded regions represent the standard deviation.
\label{fig:Skyrmion Size Effect 105mT}}\end{figure}

\paragraph{Size Dependence of Velocity and SkHE}
We examine the influence of $d_{\rm S}$, which varies over 80--180~nm (\ref{fig:Skyrmion Size Effect 105mT}a), on the $\langle v_{\rm S}\rangle$ (\ref{fig:Skyrmion Size Effect 105mT}b) and $\langle\theta_{\rm S}\rangle$ (\ref{fig:Skyrmion Size Effect 105mT}c) for $J = (5.5-5.8)\times 10^{11}$~A/m$^2$ at 105~mT. Similar to $\mu_0 H \simeq 75$~mT shown in manuscript Fig. 4d, $\langle v_{\rm S}\rangle$ is insensitive to $d_{\rm S}$. Meanwhile, the trend of $\langle\theta_{\rm S}\rangle$ with $d_{\rm S}$ is less clear across the range of $J$, largely due to inconsistencies in the outermost bins. As these outermost bins have low skyrmion counts (5\% of central bin), the size dependence analysis is improved in \ref{sec:BinningChecks} by imposing a minimum count criterion for each bin, resulting in a clearer $d_{\rm S}$ trend.

\begin{figure}[h]
\begin{centering}
\includegraphics[width=2.5in]{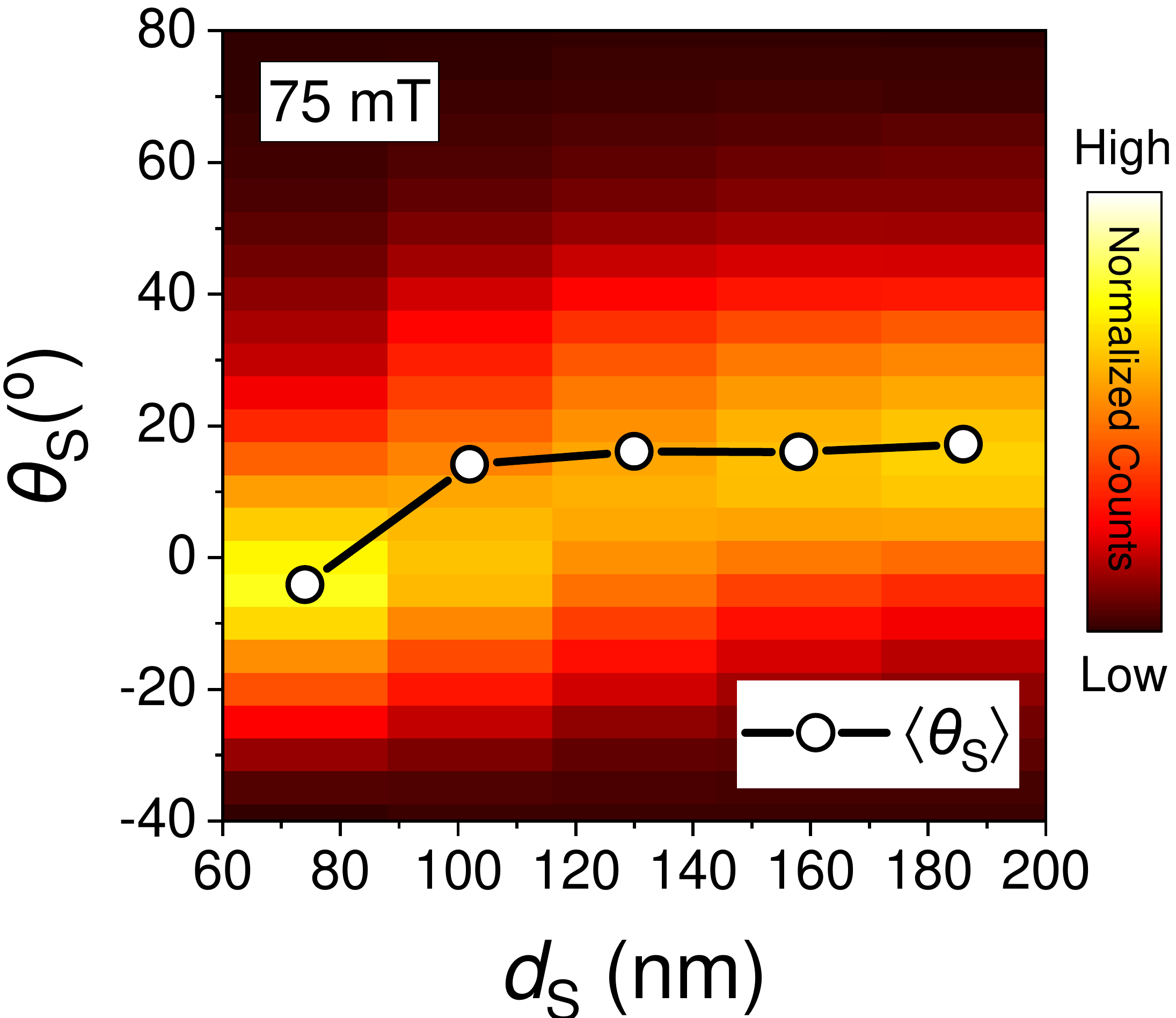}
\end{centering}
\caption[Skyrmion Size Effect at 75mT (D2)]{\textbf{Skyrmion Size Effect on Skyrmion Flow Dynamics for Second Device.} 2D histogram color plot of $\theta_{\rm S}$ against $d_{\rm S}$ for $\mu_0 H\simeq 75$~mT across all currents. The data were binned by $d_{\rm S}$. Solid markers show the average deflection, $\langle\theta_{\rm S}\rangle$, for each $d_{\rm S}$ bin.
\label{fig:Skyrmion Size Effect 75mT_d2}}\end{figure}

\paragraph{Size Dependence of SkHE for Second Device}
Here, we examine the size dependence of skyrmion dynamics in the plastic flow regime for a second device at $\mu_0 H \simeq 75$~mT and $J$ over $5.5\times10^{11}$ A/m$^2$ (\ref{fig:Skyrmion Size Effect 75mT_d2}). The skyrmions are binned by their $d_{\rm S}$ which spreads over $70-190$~nm. As shown in the 2D histogram plot, we observe a discernible increase in $\langle\theta_{\rm S}\rangle$ as $d_{\rm S}$ varies over 60--200~nm. The variation of $\langle\theta_{\rm S}\rangle$ with $d_{\rm S}$ for this device is similar in magnitude to the results presented in the manuscript Fig. 4c.

\clearpage{}

\section{Binning Effects\label{sec:BinningChecks}}

We extend the analyses carried out in manuscript Fig. 3 and 4 --
on the positional and skyrmion size dependence of skyrmion dynamics
-- with different bin parameters to verify the robustness of the
trends presented.

\begin{figure}[h]
\begin{centering}
\includegraphics[width=5.4in]{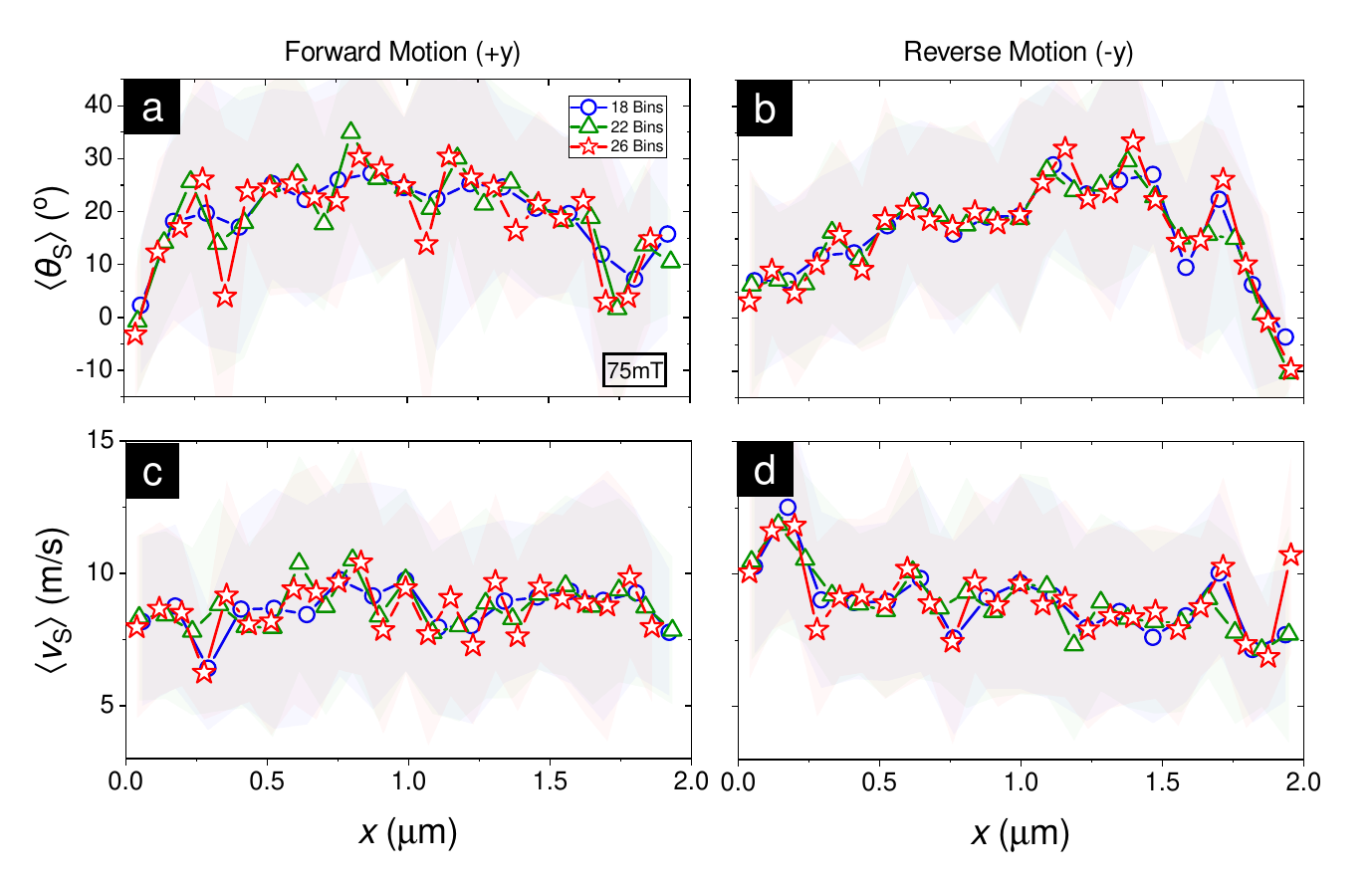}
\end{centering}
\caption[BinEdge]{\textbf{Bin Size Variation for Edge Effect Analysis. (a -- d)} Positional dependence of $\langle\theta_{\rm S}\rangle$ for the (a) forward ($J \parallel +\hat{y}$) and and (b) reverse ($J \parallel -\hat{y}$) directions, and $\langle v_{\rm S}\rangle$ for the (c) forward ($J \parallel +\hat{y}$) and (d)reverse ($J \parallel -\hat{y}$) directions in the plastic flow regime with different bin sizes. Shaded regions represent the standard deviation. \label{fig:BinEdge}}
\end{figure}

\paragraph{Edge Effect Analysis}
\ref{fig:BinEdge} shows the positional analysis (detailed in Fig. 3d, e) with bin sizes varied over a range of $\sim 40\%$ in both $J$ directions. Both the $\langle\theta_{\rm S}\rangle$ (\ref{fig:BinEdge}a, b) and $\langle v_{\rm S}\rangle$ (\ref{fig:BinEdge}c, d) dependence of position clearly display the same trend for differing bin sizes. Notably, the current direction-dependent asymmetry of $\langle\theta_{\rm S}\rangle$ (\ref{fig:BinEdge}a, b) and $\langle v_{\rm S}\rangle$ (\ref{fig:BinEdge}c,d) with respect to position were also evident, as discussed in manuscript \S D.

\begin{figure}[h]
\begin{centering}
\includegraphics[width=5.6in]{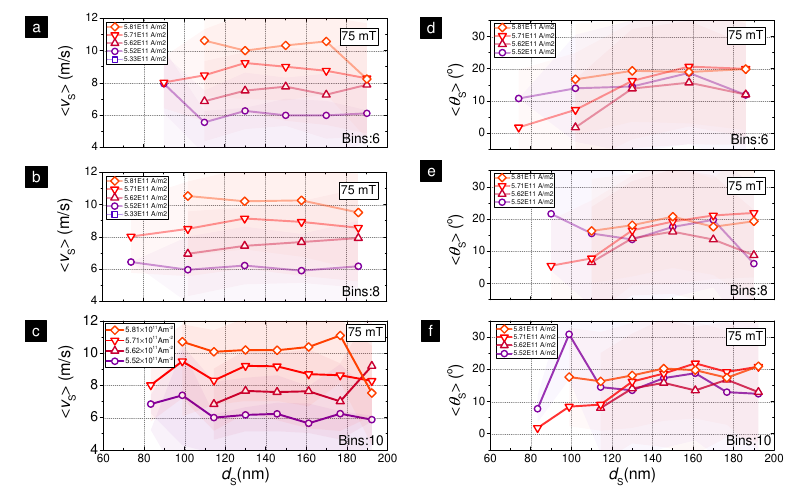}
\end{centering}
\caption[BinSkSize]{\textbf{Bin Size Variation for Skyrmion Size Effect Analysis. (a--c)} $\langle v_{\rm S}\rangle$ and \textbf{(d -- f)} $\langle\theta_{\rm S}\rangle$ in the plastic flow regime as a function of $d_{\rm S}$ for varying magnitudes of $J$ with bin sizes of (a, d) 6, (b, e) 8, and (c, f) 10 bins. Shaded regions represent the standard deviation.
\label{fig:BinSkSize}}\end{figure}

\paragraph{Skyrmion Size Effect Analysis}
The same procedure of varying the bin sizes by $\sim 40\%$ was employed in the analysis of skyrmion size effect detailed in manuscript Fig. 4d, e. The observations of the average $\langle v_{\rm S}\rangle$ steadily increasing with $J$ while relatively unchanged across $d_{\rm S}$ (Fig. 4d) are also evident here with different bin sizes (\ref{fig:BinSkSize}a -- c). Similarly, the $\langle\theta_{\rm S}\rangle - d_{\rm S}$ relationship (\ref{fig:BinSkSize}d -- f) also seems insensitive to variation in bin sizes.

\begin{figure}[h]
\begin{centering}
\includegraphics[width=2.6in]{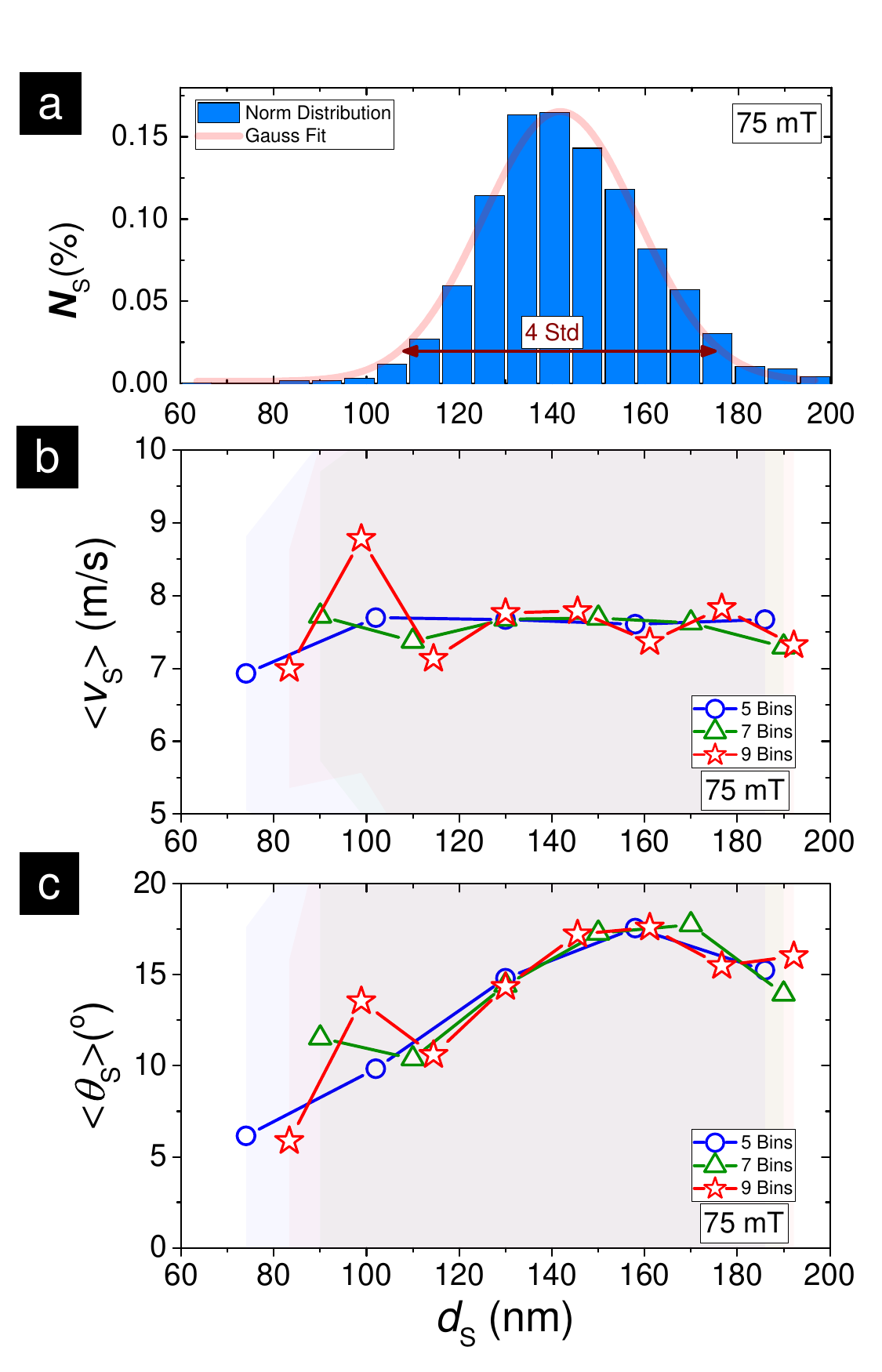}
\end{centering}
\caption[BinSkSizeFlow]{\textbf{Bin Size Variation for Size Effect Analysis Consolidated for the Plastic Flow Regime. (a)} Normalized distribution of $d_{{\rm S}}$ in the plastic flow regime fitted with a Gaussian distribution. \textbf{(b -- c)} Skyrmion size dependence of (b) $\langle v_{\rm S}\rangle$ and (c) $\langle\theta_{\rm S}\rangle$ averaged across all $J$ in the plastic flow regime with different bin sizes. Shaded regions represent the standard deviation.
\label{fig:BinSkSizeFlow}}\end{figure}

\paragraph{Skyrmion Size Effect in the Plastic Flow Regime Analysis}
The $d_{\rm S}$ dependence of $\langle v_{\rm S}\rangle$ (\ref{fig:BinSkSizeFlow}b) and $\langle\theta_{\rm S}\rangle$ (\ref{fig:BinSkSizeFlow}c) are largely reproduced within 4 standard deviations (4 Std) of the $d_{\rm S}$ distribution (\ref{fig:BinSkSizeFlow}a). Since the plastic flow dataset was further divided into different $J$ for the analysis in manuscript Fig. 4c,d (also \ref{fig:BinSkSize}a, d), it is therefore reasonable that a smaller bin number would capture more accurately the size dependence trend at the tail of the size distribution ($< 100$~ nm, \ref{fig:BinSkSizeFlow}a). To reinforce the skyrmion size dependence trends, we consolidated all the skyrmions moving in the plastic flow regime (i.e. removing the $J$ categorization) and repeated the analysis. Using different bin sizes, \ref{fig:BinSkSizeFlow}c clearly shows a monotonic increase of $\langle\theta_{\rm S}\rangle$ with $d_{\rm S}$ within 4 Std of the $d_{\rm S}$ distribution (\ref{fig:BinSkSizeFlow}a) while $\langle v_{\rm S}\rangle$ as expected is mostly constant with $d_{\rm S}$ (\ref{fig:BinSkSizeFlow}b).

\begin{figure}[h]
\begin{centering}
\includegraphics[width=5.2in]{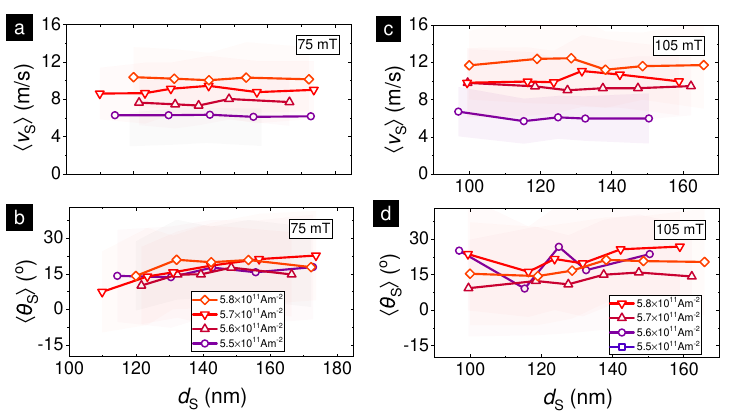}
\end{centering}
\caption[MinCtSkSizeFlow]{\textbf{Skyrmion Size Effect Analysis with Minimum Counts Criterion. (a-d)} Average velocity $\langle v_{\rm S}\rangle$ and angular deflection $\langle\theta_{\rm S}\rangle$ with a minimum of 50 skyrmion counts in each $d_{\rm S}$-bin across the plastic flow regime $(J = 5.5$ –- $5.8\times10^{11} \mathrm{A/m^2})$ at $\mu_0 H\simeq 75$~mT (a-b) and $\mu_0 H\simeq 105$~mT (c-d). Shaded regions represent the standard deviation.
\label{fig:MinCtSkSizeFlow}}\end{figure}

To improve the clarity of the skyrmion size analysis at 105 mT (\ref{fig:Skyrmion Size Effect 105mT}b-c) , we impose a minimum count of 50 for each bin. This addresses potential inconsistencies at the outermost bins that may arise from individual outliers due to low counts. For both 75 and 105 mT, the additional analysis criterion results in clearer $\langle v_{\rm S}\rangle$,  $\langle\theta_{\rm S}\rangle$ vs $d_{\rm S}$ trends. Consistent with manuscript Fig. 4, $\langle v_{\rm S}\rangle$ is insensitive (\ref{fig:MinCtSkSizeFlow}a, c)  to $d_{\rm S}$ and $\langle\theta_{\rm S}\rangle$ weakly increases with $d_{\rm S}$ (\ref{fig:MinCtSkSizeFlow}b, d).

\clearpage
\clearpage{}
\section{Micromagnetic Simulations\label{sec:SIMS}}

\begin{figure}[h]
\begin{centering}
\includegraphics[width=5in]{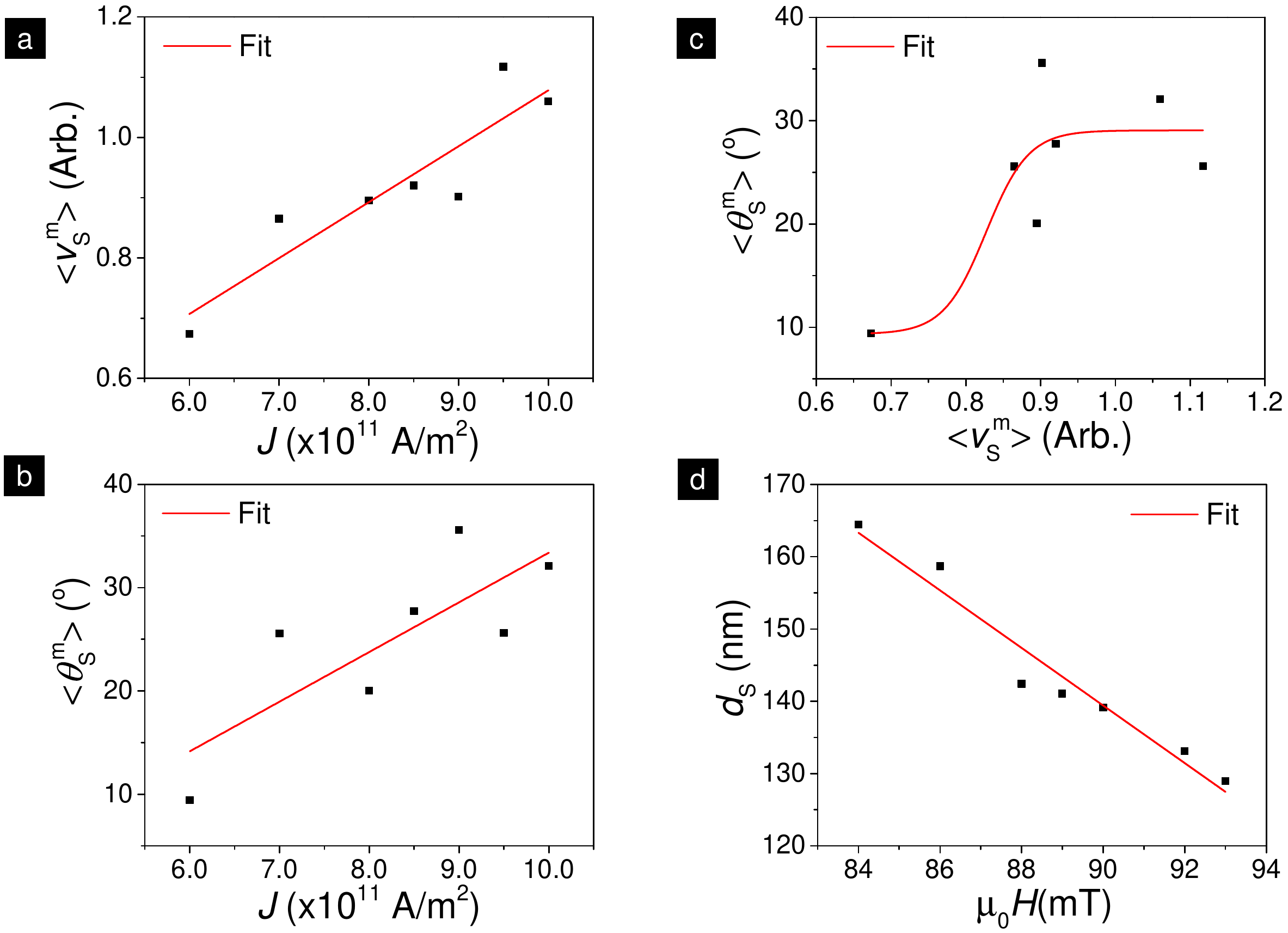}
\end{centering}
\caption[Mumag Simulations]{\textbf{Micromagnetics Simulations. (a)} Average skyrmion velocity $\langle v_{\rm S}^{\rm m}\rangle$ and \textbf{(b)} angular deflection $\langle\theta_{\rm S}^{\rm m}\rangle$ for $J$ varied from $(6-10)\times10^{11}$~A/m$^2$ at a fixed OP field of 87~mT. \textbf{(c)} $\langle\theta_{\rm S}^{\rm m}\rangle$ as a function of $\langle v_{\rm S}^{\rm m}\rangle$ showing SkHE saturation in the plastic flow regime at $J \gtrsim 9.0 \times 10^{11}$~A/m$^2$. \textbf{(d)} The variation of $d_{\rm S}$ as a function of OP field $\mu_0 H$, which was used to obtain the size dependence results. 
\label{fig:umag sim}}\end{figure}

\paragraph{Simulated Dynamics}
Our defect free micromagnetic simulations show a monotonic linear increase in the simulated skyrmion velocity, $\langle v_{\rm S}^{\rm m}\rangle$ and angular deflection, $\langle \theta_{\rm S}^{\rm m}\rangle$ (\ref{fig:umag sim}a, b) with increasing $J$. In comparison, our experiments note an exponential $\langle v_{\rm S}\rangle -J$ relationship, which additionally has a transition from the creep to plastic flow regime (manuscript Fig. 2b), with differences arising likely due to pinning effects. \ref{fig:umag sim}c shows a saturation of the $\langle\theta_{\rm S}^{\rm m}\rangle$ at $\sim 28^\circ$ at high $\langle v_{\rm S}^{\rm m}\rangle$, corresponding to $J$ larger than $9.0\times 10^{11}$~A/m$^2$, comparable to the experimental skyrmion saturation, $\theta_{\rm S}^{\rm sat} \sim 22^\circ$ in the plastic flow regime. In ensuring consistency with experimental skyrmion dynamics analysis in the plastic flow regime (manuscript Figs. 3 and 4), subsequent simulations on the $d_{\rm S}$-dependence of average $\langle v_{\rm S}^{\rm m}\rangle$ and $\langle\theta_{\rm S}^{\rm m}\rangle$ (manuscript Fig. 5) were carried out at a sufficiently high $J$ of $9.5\times 10^{11}$~A/m$^2$. \ref{fig:umag sim}d shows a linear reduction in simulated $d_{\rm S}$ by $\sim 30\%$ with increasing OP applied field. The specific range of OP fields was chosen to ensure that at most one or two sparse skyrmions were stabilized in the wire, which precludes any influence of skyrmion-skyrmion and skyrmion-stripe interactions on the simulated $\langle v_{\rm S}^{\rm m}\rangle$ and $\langle\theta_{\rm S}^{\rm m}\rangle$.

\paragraph{Comparison of Micromagnetic Simulations in Published Works}

We summarize our grain-free micromagnetic simulations and relevant granular micromagnetic simulation works, which have incorporated pinning effects on skyrmion dynamics, for the ease of comparison with our experimental results (\ref{tab:comparison-uMag}). It is worth noting that neither our grain-free micromagnetic simulations nor the granular micromagnetic simulations can fully explain our key experimental observations - the skyrmion size dependence on velocity ($v_{\rm S}$-$d_{\rm S}$) and Hall angle ($\theta_{\rm S}$-$d_{\rm S}$). This is likely because the pinning landscapes produced by the current granular implementation of disorder in micromagnetic simulations are not sufficiently representative of the true landscapes in disordered magnetic multilayers.

\begin{table}[h]
    \centering
    \begin{tabular}{ |C{5cm}|C{2cm}|C{2cm}|C{2cm}|C{2cm}|  }
        \hline
         & \boldmath{$v_{\rm S}$-$J$} &  \boldmath{$\theta_{\rm S}$-$v_{\rm S}$} & \boldmath{$v_{\rm S}$-$d_{\rm S}$} & \boldmath{$\theta_{\rm S}$-$d_{\rm S}$} \\ 
        \hline
        \textcolor{red}{\textbf{Experimental Results in Manuscript}} & \textcolor{red}{Exponential increase} & \textcolor{red}{S-curve} & \textcolor{red}{Constant} & \textcolor{red}{Weak increase} \\ 
        \hline
        \textbf{Grain Free Micromagnetic Simulations in Manuscript} & Increase & S-curve & Increase to saturation & Decrease \\ 
        \hline
        \textbf{Granular Micromagnetic Simulations in Kim, J-V. et al. \cite{Kim2017}} & Exponential increase & N.A. & N.A. & N.A. \\
        \hline
        \textbf{Granular Micromagnetic Simulations in Legrand, W. et al. \cite{Legrand2017}} & Exponential increase & N.A. & Increase to saturation & N.A. \\
        \hline
        \textbf{Granular Micromagnetic Simulations in Juge, R. et al. \cite{Juge2019}} & Exponential increase & Increase to saturation & N.A. & N.A. \\
        \hline
        \end{tabular}
    \vspace{0.5cm}
    \caption[comparison-uMag]{\textbf{Published Micromagnetic Simulations.} Comparison of the skyrmion dynamics trends reported in our experiments and grain-free micromagnetic simulations with published micromagnetic simulation results.}
    \label{tab:comparison-uMag}
\end{table}

\clearpage

\section{ Experimental Skyrmion Dynamics Characteristics in Published Works\label{sec:comparison-skDynamics}}

In \ref{tab:comparison-skDynamics}, we present a tabulated comparison of the multilayer stacks used in our work, the associated skyrmion properties, and the observed dynamic characteristics, as compared with published results on multilayer skyrmions.

\begin{table}[h]
    \centering
    \small
    \begin{adjustbox}{width=\textwidth}
        \begin{tabular}{|C{3.4cm}|C{2.5cm}|C{1.6cm}|C{1.8cm}|C{1.8cm}|C{1.7cm}|C{1.8cm}|C{1.8cm}|C{1.8cm}| }
            \hline
            \multirow{3}{3.4cm}{\textbf{Stack Composition}} & \multirow{3}{2.5cm}{\textbf{Skyrmion \newline Sizes ($d_{\rm S}$) \& \newline Densities ($n_{\rm S}$)}} & \multirow{3}{1.6cm}{\textbf{Motion Regime}} & \multirow{3}{1.8cm}{\small\textbf{Max Velocity, }\boldmath{$v_{\rm S}^{\rm{max}}$ (m/s)}} & \multirow{3}{1.8cm}{\small\textbf{Max Deflection, }\boldmath{$\theta_{\rm S}^{\rm{max}}$ ($^\circ$)}} & \multicolumn{2}{C{3.5cm}|}{\textbf{Wire Edge ($x$) Dependence}} & \multicolumn{2}{C{3.6cm}|}{\textbf{Skyrmion Size ($d_{\rm S}$) Dependence}} \\
            \cline{6-9}
            & & & & &\textbf{$v_{\rm S}$} \newline & \textbf{$\theta_{\rm S}$} \newline & \textbf{$v_{\rm S}$} \newline & \textbf{$\theta_{\rm S}$} \newline \\
            \hline
            \multicolumn{9}{|c|}{\textbf{Ferromagnetic}}\\
            \hline
            \textcolor{red}{\textbf{[Pt/Co/MgO]$_{\footnotesize 15}$}} & \textcolor{red}{$d_{\rm S}: 80\mbox{-}200\, \rm{nm}$ \newline  $n_{\rm S}: 2\mbox{-}13\, \rm{\mu m^{-2}}$} & \textcolor{red}{Stochastic \newline Creep \newline \textbf{Flow}} & \textcolor{red}{24} & \textcolor{red}{22$^\circ$ (saturated)} &\textcolor{red}{$\sim 20\%$ reduction} & \textcolor{red}{Ambipolar \footnotesize{$+22^{\circ}\rightarrow\mbox{-}5^{\circ}$}} & \textcolor{red}{Independent (flow)} & \textcolor{red}{Weak, linear (flow)} \\
            \hline
            [Pt/Co/Ir]$_{\footnotesize 10}$\cite{Legrand2017} & $d_{\rm S} \sim 100\, \rm{nm}$ \newline  $n_{\rm S} \sim 3\, \rm{\mu m^{-2}}$ & Stochastic & 1 & N.A. &- & - & - & - \\        
            \hline
            [Pt/CoFeB/MgO]$_{\footnotesize 15}$\cite{Litzius2016} & $d_{\rm S} \sim 113\, \rm{nm}$ \newline  $n_{\rm S} \sim 1\, \rm{\mu} m^{-2}$ & Creep & 105 & 32$^\circ$ & - & - & - & Inverse (creep) \\     
            \hline
            Pt/FM/Au/FM/Pt FM:Ni/Co/Ni\cite{Hrabec2017} & $d_{\rm S} \sim 150\mbox{-}200\, \rm{nm}$ \newline  $n_{\rm S} \sim 2\, \rm{\mu m^{-2}}$ & Stochastic \newline Creep & 65 & N.A. & - & - & - & - \\ 
            \hline
            [Pt/CoFeB/MgO]$_{\footnotesize 15}$\cite{Woo2016} & $d_{\rm S} \sim 200\mbox{-}300\, \rm{nm}$ \newline  $n_{\rm S} \sim 3\, \rm{\mu m^{-2}}$ & Creep & 110 & N.A. & - & - & - & - \\   
            \hline
            [Pt/Co/Ta]$_{\footnotesize 15}$\cite{Woo2016} & $d_{\rm S} \sim 200\mbox{-}300\, \rm{nm}$ \newline  $n_{\rm S} \sim 3\, \rm{\mu m^{-2}}$ & Creep & 50 & N.A. & - & - & - & - \\   
            \hline
            Ta/CoFeB/TaO$_{\footnotesize x}$\cite{Jiang2016} & $d_{\rm S} \sim 1000\, \rm{nm}$ \newline  $n_{\rm S} < 1\, \rm{\mu m^{-2}}$ & Stochastic\newline Creep \newline \textbf{Flow} & 0.75 & 35$^\circ$ & Reduction & - & - & - \\
            \hline
            Pt/Co/MgO\cite{Juge2019} & $d_{\rm S} \sim 156\, \rm{nm}$ \newline  $n_{\rm S} \sim 1.5\, \rm{\mu m^{-2}}$ & Flow & 100 & 50$^\circ$ (saturated) & - & - & - & - \\
            \hline
            [Pt/CoB/Ir]$_{\footnotesize 5}$\cite{Zeissler2020} & $d_{\rm S} \sim 400\, \rm{nm}$ \newline  $n_{\rm S} \sim 1.5\, \rm{\mu m^{-2}}$ & Creep \newline \textbf{Flow} & 6 & 9$^\circ$ (saturated) & - & - & Independent (flow) & Independent (flow) \\
            \hline
            \multicolumn{9}{|c|}{\textbf{Ferrimagnetic}}\\
            \hline
            [Pt/GdFeCo/SiN]\cite{Hirata2019} & $d_{\rm S} \sim 10,000\, \rm{nm}$ \newline  $n_{\rm S} < 1\, \rm{\mu m^{-2}}$ & Creep & N.A. & 35$^\circ$ & - & - & - & - \\
            \hline
            [Pt/GdFeCo/MgO]$_{\footnotesize 2}$\cite{Woo2018} & $d_{\rm S} \sim 180\, \rm{nm}$ \newline  $n_{\rm S} \sim 3\, \rm{\mu m^{-2}}$ & Stochastic \newline Creep \newline Flow & 50 & 25$^\circ$ (saturated)& - & - & - & - \\
            \hline
        \end{tabular}
    \end{adjustbox}
    \vspace{0.5cm}
    \caption[comparison-skDynamics2]{\textbf{Experimental Results on Skyrmion Dynamics.} Comparison of experimental results on skyrmion dynamics presented in this manuscript with published results. Key comparisons include the properties of skyrmion configurations, motion regimes, skyrmion velocity, deflection angle and dependence of dynamics on distance from the wire edge ($x$) and skyrmion size ($d_{\rm S}$).}\label{tab:comparison-skDynamics}
\end{table}
\afterpage{\clearpage}
\begin{center}
\rule[0.5ex]{0.6\columnwidth}{1pt}
\end{center}


\end{document}